\documentclass[prd, aps, superscriptaddress, showpacs]{revtex4}

\usepackage{latexsym, graphicx}

\begin{document}

\title{Entanglement creation between two causally-disconnected objects}

\author{Shih-Yuin Lin}
\email{sylin@cc.ncue.edu.tw}
\affiliation{Physics Division, National Center for Theoretical Science,
P.O. Box 2-131, Hsinchu 30013, Taiwan}
\affiliation{Department of Physics, National Cheng Kung University, 
Tainan 70101, Taiwan}
\affiliation{Department of Physics, National Changhua University of Education,
Changhua 50007, Taiwan}
\author{B. L. Hu}
\email{blhu@umd.edu}
\affiliation{Joint Quantum Institute and Maryland Center for Fundamental Physics,
University of Maryland, College Park, Maryland 20742-4111, USA}

\date{November 1, 2009}

\begin{abstract}
We study the full entanglement dynamics of two uniformly accelerated
Unruh-DeWitt detectors with no direct interaction in between but each
coupled to a common quantum field and moving back-to-back in the field 
vacuum. For two detectors initially prepared in a separable state our
exact results show that quantum entanglement between the detectors
can be created by the quantum field under some specific
circumstances, though each detector never enters the other's light
cone in this setup. In the weak coupling limit, this entanglement
creation can occur only if the initial moment is placed early enough
and the proper acceleration of the detectors is not too large or too
small compared to the natural frequency of the detectors.
Once entanglement is created it lasts only a finite 
duration, and always disappears at late times. Prior result by 
Reznik \cite{Reznik03} derived using the time-dependent perturbation
theory with extended integration domain is shown to be a limiting 
case of our exact solutions at some specific moment.
In the strong coupling and high acceleration regime,
vacuum fluctuations experienced by each detector locally always
dominate over the cross correlations between the detectors, so
entanglement between the detectors will never be generated.
\end{abstract}


\maketitle

\section{Introduction}

Quantum entanglement, the uniquely quantum mechanical feature of a
quantum system as Schr\"odinger emphasized \cite{Sc35}, is the
distinguishing resource of quantum information processing (QIP).
Its evolutionary behavior in time is thus of primary importance to the
viability and functioning of quantum computing. Many QIP candidate
systems involve qubits interacting with a quantum field, such as two
level atoms (2LA) in a cavity. Because of the intrinsically
relativistic nature of a quantum field, in that physical information
propagates at a finite speed, issues of causality are unavoidably
imbued with QIP, even though the qubits in popular schemes are likely
to move at nonrelativistic speed or remain stationary. It is of
interest to ask whether quantum entanglement can transcend such
limitations. This underlies the frequently mentioned yet
often-misconjured notion of ``quantum nonlocality" such as usually
associated with the famous EPR paper \cite{EPR}.

Following our two recent work discussing the behavior of quantum
entanglement in a relativistic setting, one with a detector (an
object with internal degrees of freedom such as a harmonic oscillator
or an atom) in relativistic motion \cite{LCH2008} and another
focusing on the relativistic features of a quantum field mediating
two inertial detectors \cite{LH2009} in this paper we study the
conditions whereby quantum entanglement between two causally
disconnected detectors (spacelike separated, outside of each other's
light cone) can be created and if so how it evolves in time. By
pushing to extreme conditions we can appreciate better this unique
feature of quantum mechanics assessed in the more complete setting of
relativistic quantum fields.

The question in focus here is, whether quantum entanglement between
two localized causally disconnected atoms can be created by the
vacuum of a common mediating quantum field they both interact with,
but not directly with each other. Refs. \cite{Reznik03, RRS05, MS06, Franson08}
affirm such a possibility whereas Refs. \cite{Braun05,LH2009} see no
such evidence. These are not contradictory claims because the setups
of the problem are not exactly the same.

Using the time-dependent perturbation theory (TDPT) with extended integration
domain, Reznik discovered that a pair of two-level atoms initially in their 
ground states will become entangled when they are uniformly accelerated 
back-to-back in the Minkowki vacuum of the field, even though in this setup 
the two atoms are causally disconnected \cite{Reznik03}. Later Massar and 
Spindel (MS) \cite{MS06} considered an exactly solvable model with two 
Unruh-Raine-Sciama-Grove (U-RSG) harmonic oscillators \cite{RSG91} moving 
in a similar fashion in (1+1) dimensions but both in equilibrium with the 
Minkowski vacuum. They discovered that entanglement can indeed be created,
but only in a finite duration of Minkowski time after the moment that the 
distance between the detectors is the minimum. 
However, contrary to Reznik's perturbative results, MS found that such 
entanglement creation process does not occur in perturbative regime. 
In our assessment, since MS were looking at the late-time steady-state 
behavior (where ``the detectors are in equilibrium with the Unruh heat bath" 
\cite{MS06}), their result cannot be compared with Reznik's TDPT result, 
which corresponds to early-time behavior of the detectors or atoms. 

To resolve the difference and understand the discrepancy our present 
investigation adopts the physical system used in \cite{LH2009} 
to a situation similar to that considered in 
\cite{Reznik03} or \cite{MS06}. 
Taking advantage of the existence of exact solutions to the model under 
study we can perform a thorough analysis and follow the system's 
evolution through their whole history 
which enables us to identify the conditions (in the motion of the two 
atoms), the approximations invoked (e.g., time dependent perturbation 
theory used in \cite{Reznik03}) and the parameter ranges where quantum
entanglement may be generated.

This paper is organized as follows. In Sec. II we introduce the
model and describe the setup in the problem. In Sec. III we derive
explicitly the expression for the cross correlators, and discuss
their evolutionary behavior. Using these correlators we examine the
exact dynamics of quantum entanglement in different conditions in
Sec. IV. We then give a comparison between the exact entanglement
dynamics and the one from the reduced density matrix (RDM) of the
truncated system. We conclude in Sec. V with some discussions.
In Appendix A we give the expression for some elements of RDM from
a first order time-dependent perturbation theory, and discuss some
subtleties in the regularization and integration domain.
In Appendix B we derive explicitly the RDM of the two truncated
detectors in an eigen-energy representation.

\section{the model}

Consider two identical, localized but spacelike separated
Unruh-DeWitt detectors, whose internal degrees of freedom $Q_A$ and
$Q_B$ are coupled to a relativistic massless quantum scalar field as
described in \cite{LH2009}, undergoing uniform acceleration in
opposite directions as described in \cite{Reznik03}. The action is
given by Eq.(1) in \cite{LH2009}, but now the trajectories of the
detectors are chosen as $z_A^\mu = (a^{-1}\sinh a\tau, a^{-1}\cosh
a\tau,0,0)$, and $z_B^\mu = (a^{-1}\sinh a\tau, -a^{-1}\cosh
a\tau,0,0)$, parametrized by their proper times $\tau$ and proper
acceleration $a$. In this setup the detectors never enter into the
light cone of each other, so they cannot exchange classical
information and energy.

Suppose the initial state of the combined system is a direct product
of the Minkowski vacuum of the field and a separable state of the
detectors in the form of a product of the Gaussian states with minimum
uncertainty for each free detector, represented by the Wigner function
\begin{equation}
  \rho(Q_A,P_A,Q_B,P_B) = {1\over \pi^2\hbar^2}\exp -{1\over\hbar}
      \left(\alpha^2 Q_A ^2 +\alpha^{-2}P_A^2 
     +\beta^2 Q_B^2 +\beta^{-2}P_B^2\right), 
\label{initGauss}
\end{equation}
where $P_A$ and $P_B$ are the conjugate momenta of $Q_A$ and $Q_B$,
respectively, and $\ln (\alpha^2/\Omega)$ and 
$\ln(\beta^2/\Omega)$ are the squeeze parameters.
Then the quantum state of the combined system is always
Gaussian during the evolution since the action is quadratic. In this
case the separability of the two detectors can be well defined by the
covariance matrix of the detectors throughout their history.

We study the entanglement dynamics by calculating the exact evolution
of the two-point functions or correlators, as in \cite{LH2009}. To
compare with the results in \cite{Reznik03} in the time-dependent
perturbation theory 
regime, we will show the reduced density
matrix (RDM) of the two detectors in an eigen-energy representation,
then truncate the energy spectra of the detectors to include only
their ground states and first excited states so that they look like
two two-level systems. Using the explicit expressions for the
elements of the RDM of the truncated system we can compare the criteria
of separability in the dynamics of the exact and truncated systems.

\section{cross correlators}

When the detectors are in a Gaussian state the RDM obtained from
integrating over the quantum field is fully determined by the two-point
functions or correlators of the detectors. If we could obtain the time
evolution of each correlator of the detectors, we have the full
dynamics of this model.

By virtue of the factorized initial state for the combined system and
the linear interaction, each correlator splits into two parts as
$\left<\cdots\right>=\left<\cdots\right>_{\rm a}+ \left<\cdots
\right>_{\rm v}$. The a-part describes the evolution of the initial
zero-point fluctuation in the detector, while the v-part accounts for
the response to the vacuum fluctuations of the field. Since the quantum
field effects such as retardation from one detector will never reach
the other in the setup of this paper, no higher order correction from
mutual influences is needed. So in this setup the expressions for the
self correlators of a single detector in \cite{LH2006} and \cite{LH2009}
are actually exact, and the v-part of the self correlators there can be
directly applied here with $\left<\right. Q_A(\eta)^2\left.\right>_{
\rm v}=\left<\right.Q_B(\eta)^2 \left.\right>_{\rm v}$, $\left<\right.
P_A(\eta)^2\left.\right>_{\rm v}= \left<\right. P_B(\eta)^2\left.
\right>_{\rm v}$, and $\left<\right.Q_A(\eta),P_A(\eta)\left.\right>_{
\rm v}=\left<\right.Q_B(\eta),P_B(\eta)\left.\right>_{\rm v}$. The
a-part of the self correlators for $Q_A$ and $Q_B$ could be different
if we take different values of $\alpha$ and $\beta$ in the initial state
$(\ref{initGauss})$, but the calculation is still straightforward. The
remaining task is to calculate the cross correlators between the two
detectors.

The a-part of the cross correlators always vanishes
since the initial state $(\ref{initGauss})$ is factorizable for
detectors $Q_A$ and $Q_B$ at the initial moment and no retarded
mutual influence between them ever arises after the coupling is
switched on. So the cross correlator $\left<\right. Q_A(\eta),
Q_B(\eta')\left.\right>$ contains contributions solely from the
v-part, or vacuum fluctuations of the field. Explicitly, it is
obtained by performing the following two-dimensional integration,
\begin{equation}
  \left<\right. Q_A(\eta),Q_B(\eta')\left.\right> = {\lambda_0^2
    \over \Omega^2}{\rm Re}\,\int_{\tau_0}^\tau ds
    \int_{\tau'_0}^{\tau'} ds' e^{-\gamma(\tau - s)-\gamma(\tau'-s')}
    \sin\Omega(\tau-s)\sin\Omega(\tau'-s') D^+(z_A^\mu(s), z_B^\nu(s')),
\label{QXdef}
\end{equation}
where $\lambda_0$ is the coupling constant between each detector and
the field, $\gamma\equiv \lambda_0^2/8\pi$, $\Omega$ is the natural
frequency of each detector, $\eta\equiv \tau-\tau_0$, $\eta'\equiv
\tau'-\tau_0'$ are the durations of interaction from the initial
moments $\tau_0$ and $\tau'_0$ when the coupling with the field is
switched on to the proper times $\tau$ and $\tau'$ of detectors $Q_A$
and $Q_B$, respectively, and $D^+(z_A^\mu(s), z_B^\nu(s'))$ is the
positive frequency Wightman function of the massless scalar field,
Eq. $(\ref{DpAB})$. In Appendix \ref{TDPTapx}, we learned that one
should take the value of the mathematical cutoff $\epsilon$ in the
expression $(\ref{DpAB})$ to be zero in calculating the cross
correlators. Then the width of the $|D^+|$ ridge on $(s, s')$ plane
(see Appendix \ref{DpRidge}) is infinity in the $\Delta$
direction, and $\left<\right. Q_A(\eta), Q_B(\eta')\left.\right>$ can
be written in the closed form,
\begin{eqnarray}
  & & \,\,\,\left<\right. Q_A(\eta),Q_B(\eta')\left.\right>
    = {\hbar\gamma\over \pi \Omega^2} {\rm Re}\, \left\{
    \left[ {i\gamma\over\Omega}-(\Omega+i\gamma)\partial_\Omega\right]
    F_{K_-}\left(-e^{a(\tau+\tau')}\right) \right. \nonumber\\ & & +
     e^{-\gamma\eta'}\left[-{i\gamma\over \Omega}\cos\Omega\eta' +
     i\sin\Omega\eta'+ e^{i\Omega\eta'}(\Omega+i\gamma)
     \partial_\Omega\right]
    F_{K_-}\left(-e^{a(\tau+\tau'_0)}\right) \nonumber\\ & & +
     e^{-\gamma\eta}\left[-{i\gamma\over \Omega}\cos\Omega\eta +
     i\sin\Omega\eta + e^{i\Omega\eta}(\Omega+i\gamma)
     \partial_\Omega\right]
    F_{K_-}\left(-e^{a(\tau'+\tau_0)}\right) \nonumber\\ & & +\left.
     e^{-\gamma(\eta+\eta')}\left[ -e^{i\Omega(\eta+\eta')}
      \left(1+(\Omega+i\gamma)\partial_\Omega\right)+
      \left({i\gamma\over\Omega}+1\right)\cos\Omega(\eta-\eta')\right]
    F_{K_-}\left(-e^{a(\tau_0+\tau'_0)}\right) \right\},
\label{exactQX}
\end{eqnarray}
where  $K_- \equiv (\gamma -i\Omega)/a$, and $F_K(x) \equiv -
{}_2F_1 (1, 1+K, 2+K, x)\times x/(1+K)$. If $\tau_0$, $\tau_0' \to -\infty$ and       
$\gamma\eta$, $\gamma\eta' \to \infty$ while $a\tau$, $a\tau'$ are finite,      %
only the first line of $(\ref{exactQX})$ which is a function of $\tau+
\tau'$ will survive. This is the counterpart of the equilibrium result          %
obtained by Massar and Spindel in the U-RSG model \cite{MS06}.                

Other cross correlators are obtained straightforwardly from $\left<\right. 
Q_A(\eta), Q_B(\eta')\left.\right>$ by proper-time derivatives, 
for example, $\left<\right.Q_A(\eta),P_B(\eta')\left.\right> =  
\partial_{\tau'}\left<\right. Q_A(\eta),Q_B(\eta')\left.\right>$, 
and so on.

Below we consider the case with $\tau'=\tau$ and $\tau_0'=\tau_0$. In
this case the time-slicing scheme is equivalent to Minkowski times. Given
\begin{equation}
  F_K(-1/z) \to  {1\over K}-
    {\pi z^K\over \sin \pi K}-{z\over K-1}+{z^2\over K-2} + O(z^3)
\end{equation}
as $z \to 0^+$, one can see that $\left<\right.Q_A(\eta),Q_B(\eta')\left.
\right>$ always vanishes as $a\tau$, $a(\tau+\tau_0)$, and $\gamma\eta
\to \infty$. Thus the detectors must be separable at late times. But the
transient behavior of the cross correlators could be more complicated.
In particular, in the regime with $\gamma \ll \Omega, a$ and $\gamma
(\tau-\tau_0) \ll 1$, but $a(\tau-\tau_0)\gg 1$, one can see that the
cross correlators manifest the following multi-stage behaviors.

\begin{figure}
\includegraphics[width=8cm]{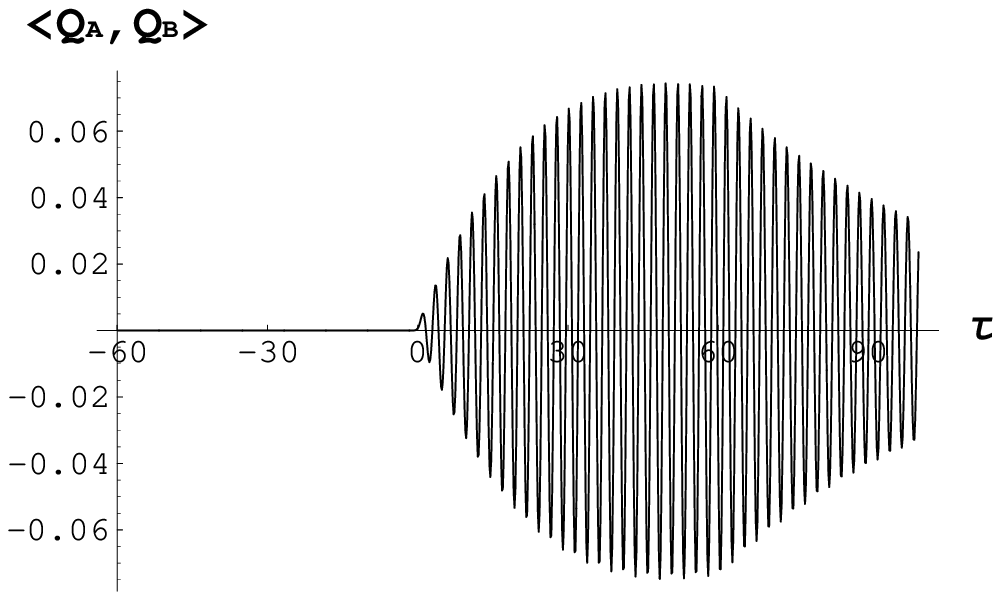}
\includegraphics[width=8cm]{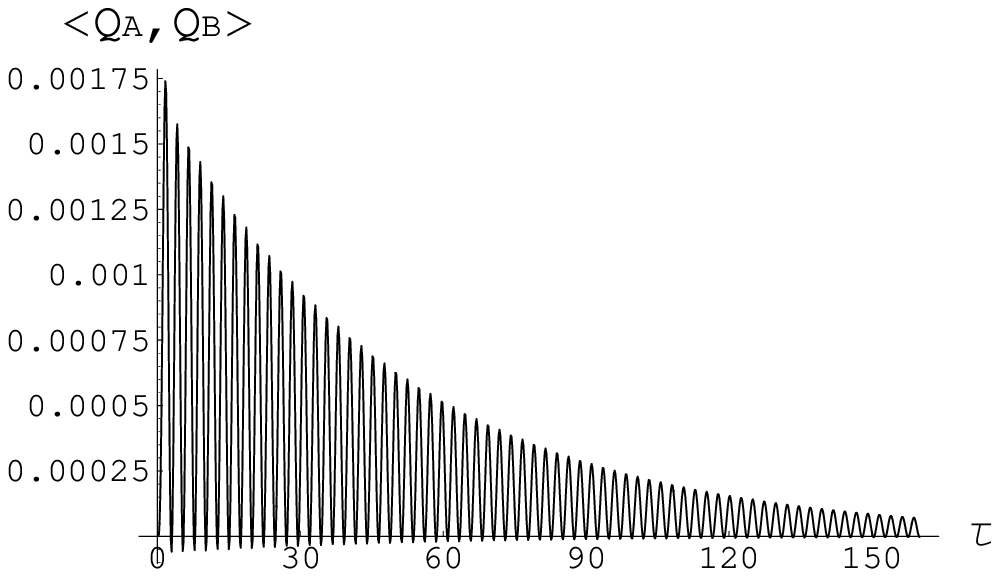}
\caption{Evolution of $\left<\right. Q_A,Q_B\left.\right>$ for the
detectors initially in their ground state, with $\gamma=0.01$,
$\Omega=1.3$, $a=2$, $\hbar=1$,
and starting at $\tau_0= -60$ (left) and $\tau_0= 0$ (right). In the
left plot one can see that when $0\alt\tau\alt -\tau_0=60$,
$\left<\right. Q_A,Q_B\left.\right>$ behaves according to
$(\ref{t0ggt})$, while  after $\tau > 60$ it behaves according to
$(\ref{tggt0})$. In the right plot one sees that for $\tau_0=0$,
$\left<\right. Q_A,Q_B\left. \right>$ behaves like $(\ref{zerot0})$
and its amplitude never grows.} \label{QAQBdemo}
\end{figure}

In the cases with large $-\tau_0 > 0$ but still $\gamma |\tau_0| \ll
1$, $\left<\right. Q_A,Q_B\left.\right>$ behaves differently in three
stages respectively, as illustrated in FIG. \ref{QAQBdemo}(left):

(i) When $\tau<0$, the value of $\left<\right. Q_A,Q_B\left.\right>$
is extremely small though exponentially growing ($\sim e^{2a\tau}$) if
$|a|$ is not very small.

(ii) When $1 \alt a\tau < -a \tau_0$, the first line of $(\ref{exactQX})$
dominates, so $\left<\right. Q_A,Q_B\left.\right>$ oscillates like
\begin{equation}
  \left<\right. Q_A,Q_B\left.\right> \approx
  {\hbar\gamma e^{-2\gamma\tau}\over\Omega \sinh{\pi\Omega\over a}}
  \left[-2\tau \cos 2\Omega\tau + {\pi\over a}\coth{\pi\Omega\over a}
  \sin 2\Omega\tau\right], \label{t0ggt}
\end{equation}
where the $\tau\cos 2\Omega\tau$ term dominates at large $\tau$, when
the amplitude of the oscillation grows almost linearly if $\gamma\tau\ll
1$. The amplitude of the oscillating $\left<\right. Q_A,Q_B\left.\right>$
will reach the maximum at $\tau \approx |\tau_0|$ if $|\tau_0| <
1/2\gamma$ with the maximum amplitude about
\begin{equation}
  {2\hbar\gamma\tau_0 e^{2\gamma\tau_0}\over\Omega \sinh{\pi\Omega\over a}},
\label{maxQxtau0}
\end{equation}
otherwise at $\tau \approx 1/2\gamma$ with the maximum amplitude about
\begin{equation}
  {\hbar\over e\Omega\sinh {\pi\Omega\over a}}.
\label{maxQx}
\end{equation}

(iii) After $\tau > -\tau_0$, the second and the third lines of
$(\ref{exactQX})$ become important and cancel the $\tau\cos 2\Omega\tau$
behavior, so one has
\begin{equation}
  \left<\right. Q_A,Q_B\left.\right> \approx
  {\hbar\gamma e^{-2\gamma\tau}\over \Omega^2 \sinh{\pi\Omega\over a}}
  \left[ 2 \Omega \tau_0\cos 2\Omega\tau -
    {\pi\Omega\over a}\coth{\pi\Omega\over a}\sin 2\Omega\tau
    + 2\sin\Omega(\tau-\tau_0)\cos\Omega(\tau+\tau_0)\right],
\label{tggt0}
\end{equation}
which oscillates with slowly decaying amplitude ($\approx 2\hbar\gamma
e^{-2\gamma\tau}|\tau_0| /\Omega \sinh(\pi\Omega/a)$
if $\Omega|\tau_0|\gg 1$).

The underlying reason for the above three-stage behavior is similar to
the one for explaining the behavior of $\rho_{11,00}^R$, which is
discussed below $(\ref{DpAB})$. 
Later we will see that this three-stage profile of $\left<\right.
Q_A,Q_B\left.\right>$ will be present in the entanglement dynamics in
the same regime.

Two special cases are worthy of mention here. First, if $\gamma\tau_0
\to -\infty$, $\left<\right.Q_A,Q_B\left.\right>$ will never enter stage
(iii). It will always behave like $(\ref{t0ggt})$ at large positive $\tau$.
Second, if $\tau_0=0$, $\left<\right. Q_A,Q_B\left.\right>$ has no stages
(i) and (ii). Rather, it starts with stage (iii) and
\begin{eqnarray}
  & &\left<\right. Q_A,Q_B\left.\right>\approx {\hbar\gamma\over\pi\Omega^2}
  e^{-2\gamma\tau}\left\{ {\pi\over 2\sinh{\pi\Omega\over a}} \left( 1 -
    {\pi\Omega\over a}\coth{\pi \Omega\over a}\right)\sin 2\Omega\tau
  + \right. \nonumber\\ & & \left. {\Omega\over 4a}{\rm Re}\,\left[
    i\psi^{(1)}\left(-{i\Omega\over 2a}\right)-i \psi^{(1)}\left({1\over 2}
    -{i\Omega\over 2a}\right)\right]\cos 2\Omega\tau
  + {1\over 2}{\rm Re}\, \left[ \psi\left( -{i\Omega\over 2a}\right)-
    \psi\left({1\over 2}-{i\Omega\over 2a}\right)\right]
    (1-\cos 2\Omega\tau)\right\},
\label{zerot0}
\end{eqnarray}
which oscillates in proper time $\tau$ with a very small and slowly
decaying amplitude, as shown in FIG. \ref{QAQBdemo}(right).

Note that the behavior of cross correlators is non-Markovian because
it depends on the fiducial time, namely, the initial moment $\tau_0$.
So is the entanglement dynamics.

\section{entanglement dynamics}

\subsection{full dynamics}

The degree of quantum entanglement between the two detectors in
Gaussian states can be measured by the values of the logarithmic
negativity $E_{\cal N}$ or the quantity $\Sigma$ defined
in \cite{LH2009}.
To demonstrate how the evolution of the cross correlators affect the
entanglement dynamics, however, we calculate below $c_-$ defined by 
\begin{equation}
  c_\pm \equiv \left[Z \pm \sqrt{Z^2-4\det {\bf V}}\over 2
    \right]^{1/2}
\label{SympSpec}
\end{equation}
with the covariance matrix $V$ in $(\ref{CovarMtx})$ and
\begin{equation}
  Z = \det {\bf v}_{AA} + \det {\bf v}_{BB} - 2 \det {\bf v}_{AB}.
\end{equation}
$c_-$ ($c_+$) is the smaller (larger) eigenvalue in the symplectic
spectrum of the partially transposed covariance matrix plus a
symplectic matrix \cite{LH2009}. If $c_- < \hbar/2$, the two
detectors are entangled, otherwise separable. From $c_\pm$ one can
easily determine the values for the quantity $\Sigma =[c_+^2-(\hbar^2/
4)][c_-^2-(\hbar^2/4)]$ we used before, and the logarithmic negativity
by $E_{\cal N}= \max\{0, -\log_2(2c_-/\hbar)\}$, where the information
in the range $c_- > \hbar/2$ is beyond reach.

\begin{figure}
\includegraphics[width=8cm]{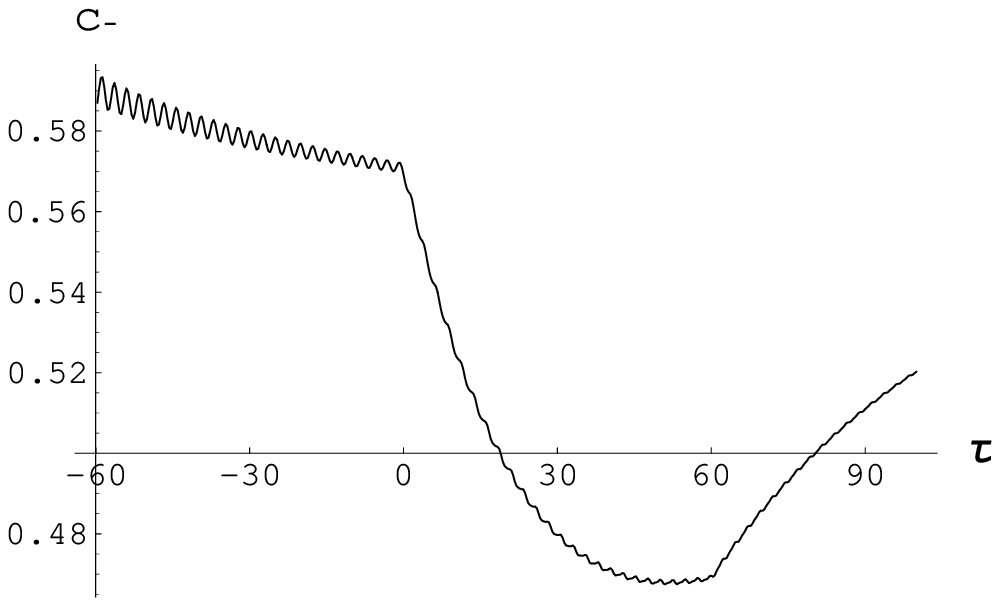}
\includegraphics[width=8cm]{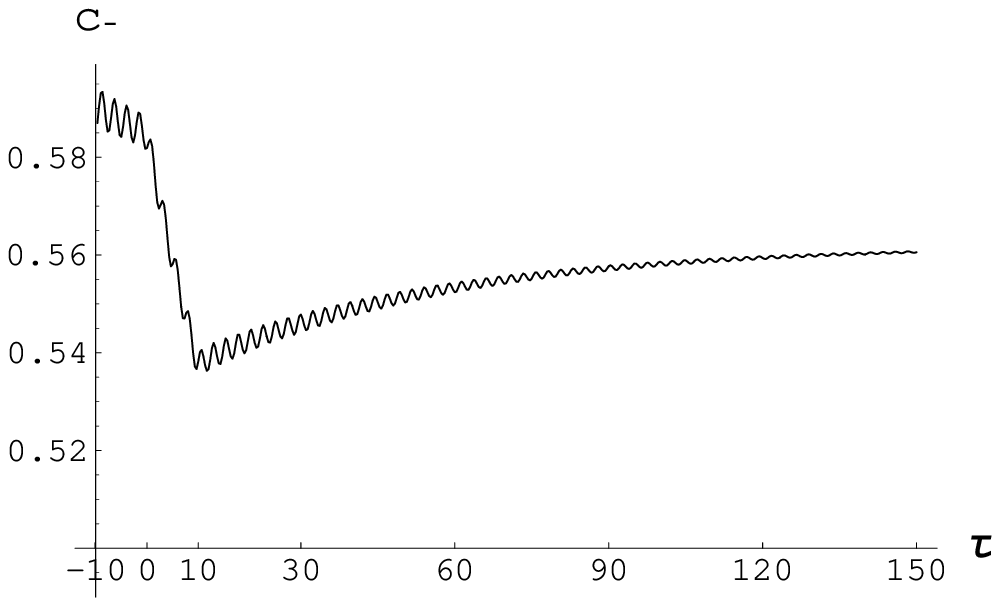}
\caption{Evolution of $c_-$ of the detectors initially in their
ground states ($\alpha=\beta=\sqrt{\Omega_r}$). (Left) The
parameters in the left plot are the same as those in Fig.
\ref{QAQBdemo}(left) (i.e., $\tau_0 = -60$, etc.) to facilitate
direct comparison. One can see that the three-stage profile of
$\left<\right.Q_A,Q_B\left. \right>$ emerges in the evolution of
$c_-$, and creates transient entanglement during $19 \alt \tau \alt
80$, which is much longer than the natural period of the detectors.
(Right) The right plot assumes the same conditions as the left plot
except that the initial moment is $\tau_0 = -10$. Now $c_-$ is always
greater than $0.5$ because the cross correlators have not had
sufficient time to become strong enough to create quantum
entanglement. This shows clearly that the entanglement creation
process is intrinsically non-Markovian in nature.}
\label{exactDyn001}
\end{figure}

Let us first consider the case with both detectors initially in their
ground states, namely, $\alpha =\beta= \sqrt{\Omega_r}$, where
$\Omega_r \equiv \sqrt{\Omega^2+\gamma^2}$ is the renormalized
natural frequency of the detectors. In FIG. \ref{exactDyn001} (left)
we can see that the three-stage profile of
$\left<\right.Q_A,Q_B\left. \right>$ in FIG. \ref{QAQBdemo}(left)
emerges in the evolution of $c_-$. A transient entanglement is
created as the amplitude of the cross correlators grows, then
decreases as the amplitude of the cross correlators decays, and
totally disappears at a finite time. The created entanglement could
remain in a duration much longer than the natural period of each
detector.

In the ultraweak coupling limit ($\gamma\Lambda_1 \ll a,\Omega$) the
feature of the cross correlators is even clearer in the entanglement
dynamics. Indeed, in this limit we have \cite{LH2006, LH07a}
\begin{equation}
  \left<\right.Q^2_A\left.\right> \approx {\cal Q} + O(\gamma\Lambda_0),
  \label{Q2weak}
\end{equation}
\begin{equation}
    \left<\right.P^2_A\left.\right> = \left<\right.P^2_B\left.\right>
    \approx\Omega^2 {\cal Q} + {2\over\pi}\gamma\hbar
     \left(\Lambda_1-\ln {a\over\Omega}\right)
    + O(\gamma\Lambda_0),
  \label{P2weak}
\end{equation}
with the constants $\Lambda_0$ and $\Lambda_1$ corresponding to the
time scale of switching on the interaction and the time resolution of
the detector, respectively, and
\begin{equation}
 {\cal Q}\equiv{\hbar\over 2\Omega}\left[ e^{-2\gamma\eta}+
    \coth{\pi\Omega\over a}\left(1-e^{-2\gamma\eta}\right)\right],
\end{equation}
while $\left<\right.P_A, Q_A\left.\right>$ and
$\left<\right.P_B,Q_B\left.\right>$ are $O(\gamma)$ and negligible.
When $\tau$ is large, one can write the cross correlators as
$\left<\right.Q_A, Q_B\left.\right> \approx \chi \cos 2\Omega\tau$,
$\left<\right.Q_A, P_B\left.\right> = \left<\right.P_A, Q_B\left.\right>
\approx -\Omega \chi \sin 2\Omega\tau$, and $\left<\right.P_A, P_B\left.
\right> \approx -\Omega^2 \chi \cos 2\Omega\tau$, where
\begin{equation}
  \chi \equiv -{2\hbar\gamma e^{-2\gamma\tau} \over\Omega\sinh
    {\pi\Omega\over a}}\left[ \tau \theta(\tau) -
    (\tau + \tau_0)\theta(\tau+\tau_0)\right]
  \label{chiweak}
\end{equation}
is the envelop of the oscillating cross correlators. Then one has
\begin{equation}
  c_- \approx \Omega\left( {\cal Q}-\left| \chi\right|\right)
    + O(\gamma \Lambda_0, \gamma\Lambda_1),
\label{cmwc}
\end{equation}
which is less than $\hbar/2$ if the detectors are entangled. Now one
can easily see how the profile of the cross correlators ($\sim
|\chi|$) enter in the entanglement dynamics.

\begin{figure}
\includegraphics[width=8cm]{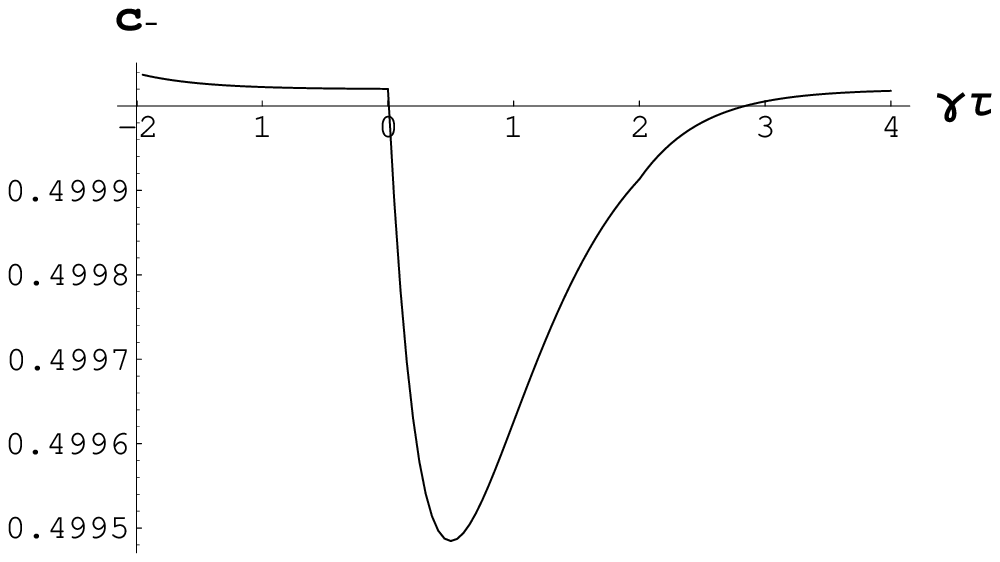}
\includegraphics[width=8cm]{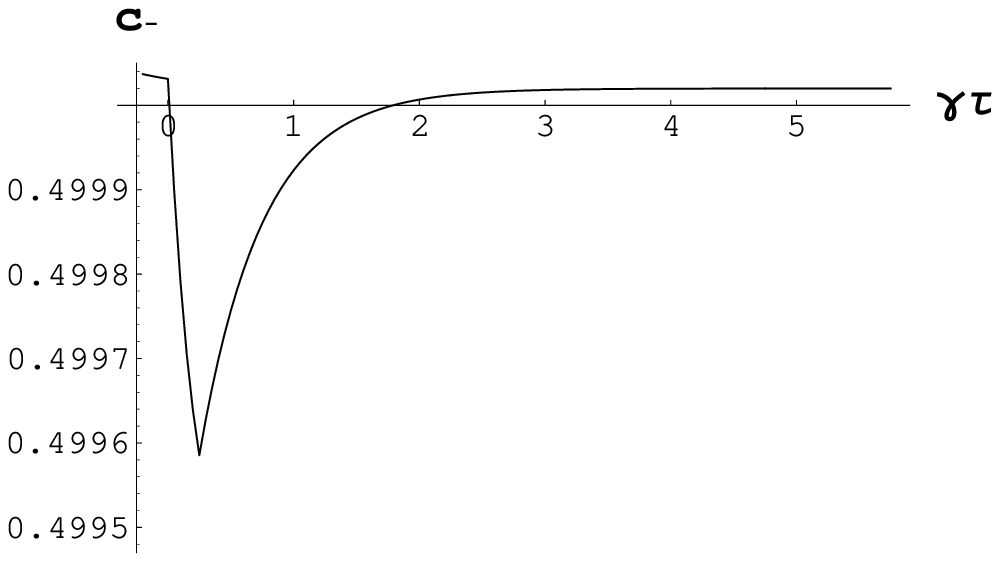}
\caption{Evolution of $c_-$ of the detectors initially in their
ground state ($\alpha=\beta=\sqrt{\Omega_r}$) in the ultraweak
coupling limit, with $\gamma=10^{-5}$, $\Omega=2.3$, $\hbar=a=1$,
$\Lambda_0=\Lambda_1=20$, $\tau_0= -2/\gamma$ (left) and
$\tau_0=-1/4\gamma$ (right). Quantum entanglement is created ($c_- <
\hbar/2$) after $\tau \agt 0$, but disappears $(c_- \ge \hbar/2)$ at
late times.} \label{exactDynWeak}
\end{figure}

If $|\tau_0|\gg 1/\gamma$, the analysis is the simplest: When
$\tau\agt 0$, ${\cal Q}$ has been in its late-time constant value
and $\Lambda_0$ term decays away. Then one has
\begin{equation}
  c_-\approx\Omega\left( {\hbar\over 2\Omega}\coth{\pi\Omega\over a}
    -\left| \chi\right|\right) + {\gamma\hbar\over\pi\Omega}\left(
    \Lambda_1 -\ln{a\over\Omega}\right),
  \label{cmwcinf}
\end{equation}
which, together with ($\ref{maxQx}$) and $(\ref{Q2weak})$, imply that
there will be transient entanglement creation outside the light cone
after $\tau\agt 0$ if
\begin{equation}
  {\pi\over \ln (2\gamma\Omega/e\gamma\Lambda_1)}  \alt
  {a\over\Omega} \alt  {\pi\over \ln (e/2)} \approx 10.238,
\label{ECuplim}
\end{equation}
for the cases with $\Lambda_1 \gg |\ln [\pi/\ln (2\gamma\Omega/e\gamma
\Lambda_1)]|$ in the ultraweak coupling limit. That is, to generate
entanglement, the proper acceleration of the detectors or the Unruh
temperature experienced by the detectors cannot be too small or too
large, otherwise the self correlators will always dominate and the
entanglement will never be created in this case. For the cases with
the value of $a/\Omega$ satisfying $(\ref{ECuplim})$, by substituting
$(\ref{chiweak})$ into $(\ref{cmwcinf})$, one can further estimate the
moment of entanglement creation $\tau_E\approx-W_0(-\zeta)/2\gamma$
and the disentanglement time $\tau_{dE}\approx-W_{-1}(-\zeta)/2\gamma$,
where
\begin{equation}
  \zeta \equiv {e^{-\pi\Omega/a}\over 2} +{\gamma\over\pi\Omega}\left(
   \Lambda_1 - \ln{a\over\Omega}\right)\sinh {\pi\Omega\over a},
\end{equation}
$W_k$ is the $k$-th product log or Lambert $W$ function, which is the
inverse function of $f(W) = W\exp W$. For example, for $\tau_0\to-\infty$
with other parameters the same as those in Fig. \ref{exactDynWeak},
one has $\tau_E\approx 1\times 10^3$ and $\tau_{dE}\approx 2.8\times 10^5$.
Note that in this setup the moment of entanglement creation $\tau_E$
is always positive, when the two detectors are moving apart.
While the $\gamma\Lambda_1$ term in $(\ref{cmwcinf})$ is small
compared to the ${\cal Q}-|\chi|$ term, it can strongly affect the
values of $\tau_E$ and $\tau_{dE}$ if $e^{-\pi\Omega/a}$ is very small,
{\it i.e.}, $\Omega({\cal Q}-|\chi|)$ is very close to $\hbar/2$.

For the cases with a smaller $|\tau_0|\sim O(1/2\gamma)$ the situation is 
similar. Different values of $\tau_0$ would give about the same $\tau_E$, 
if entanglement creation still happens, while the disentanglement time
$\tau_{dE}$ and the minimum value of $c_-$ (thus the upper limit of $a/
\Omega$ for entanglement creation) can be quite different but of the same
order as those with $|\tau_0|\gg 1/\gamma$. An example of entanglement
creation in ultraweak coupling limit is given in Fig. \ref{exactDynWeak}.

When $a$ gets smaller than $\gamma$, the above approximation in ultraweak
coupling limit fails \cite{LH2006}. The entanglement dynamics for the
cases with both detectors at rest ($a=0$) has been discussed in Ref.
\cite{LH2009}, while the separation $d$ there should be taken as $2/a
\gg 1$ in this setup.

\begin{figure}
\includegraphics[width=8cm]{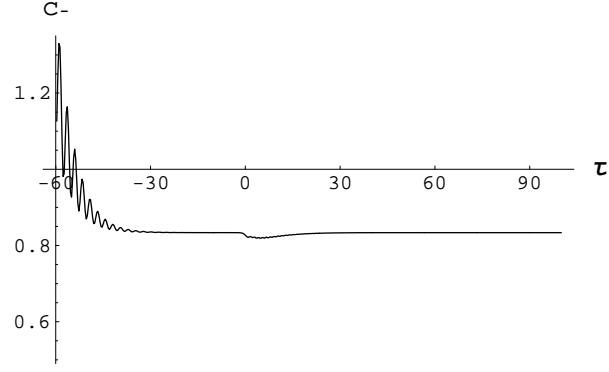}
\caption{Evolution of $c_-$ of the detectors initially in their
ground states in the strong coupling limit, with $\gamma=0.1$,
$\Omega=1.3$, $\hbar=a=1$, $\Lambda_0=\Lambda_1=20$, and $\tau_0=
-60$. Owing to the strong coupling the self correlators always
dominate over the cross correlators , so quantum entanglement is
never created.} \label{Dynstrong}
\end{figure}

In the strong coupling limit, the self correlators always dominate
over the cross correlators, due to the manifestation of the
long-range autocorrelation in the field that each detector
experienced locally in space. So quantum entanglement is never
created if the coupling is sufficiently strong (see Fig.
\ref{Dynstrong}).

\begin{figure}
\includegraphics[width=4cm]{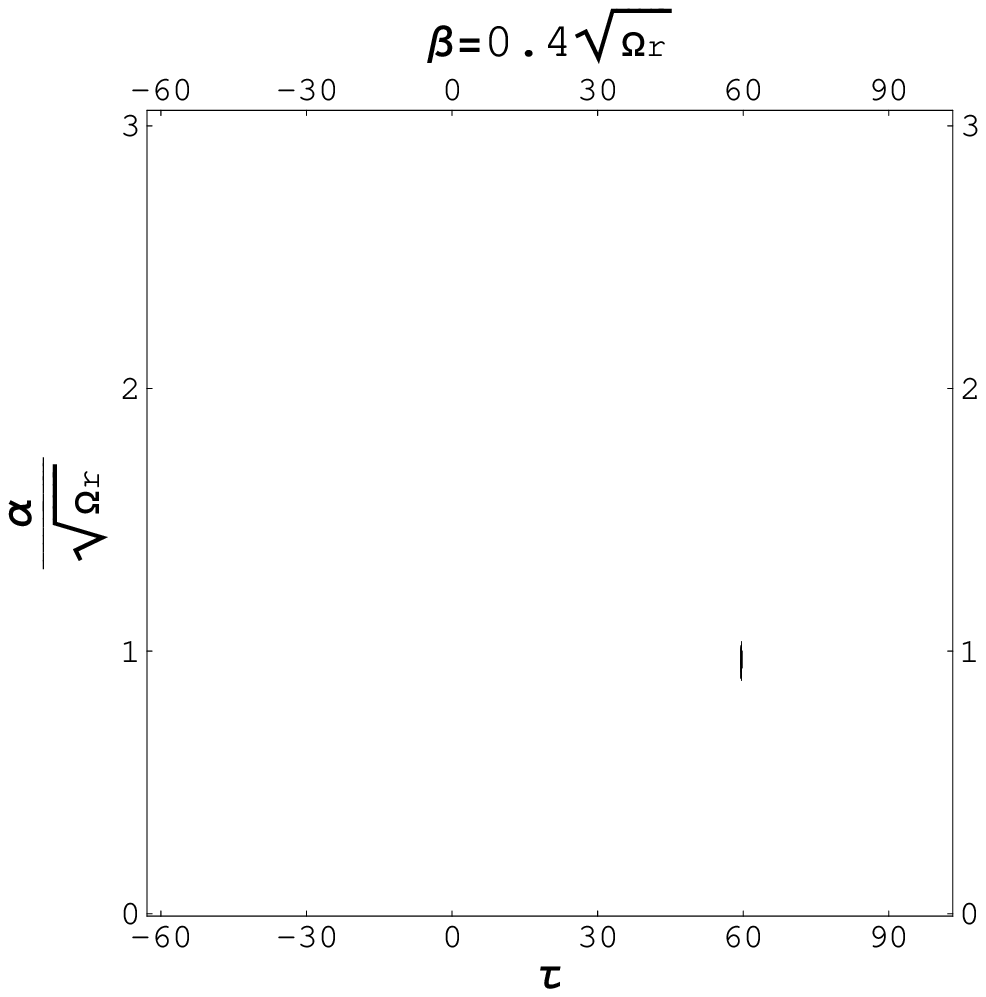}
\includegraphics[width=4cm]{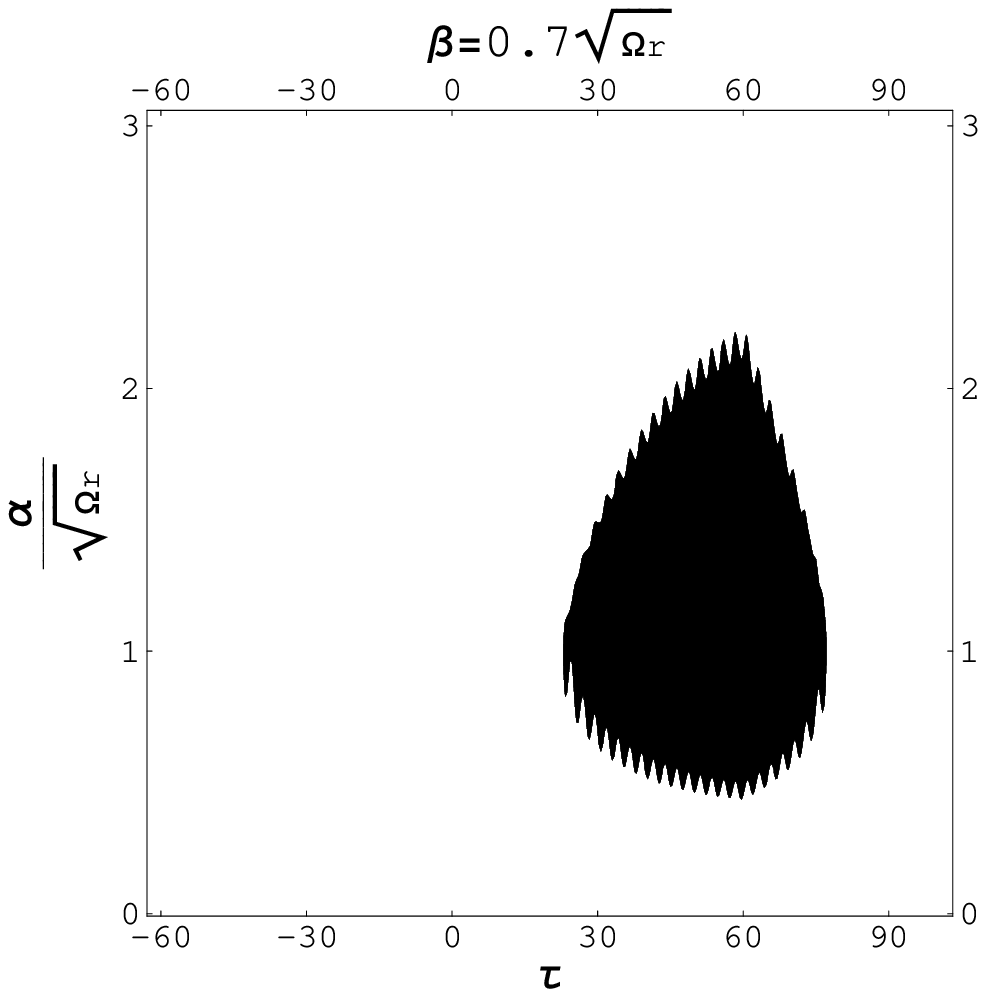}
\includegraphics[width=4cm]{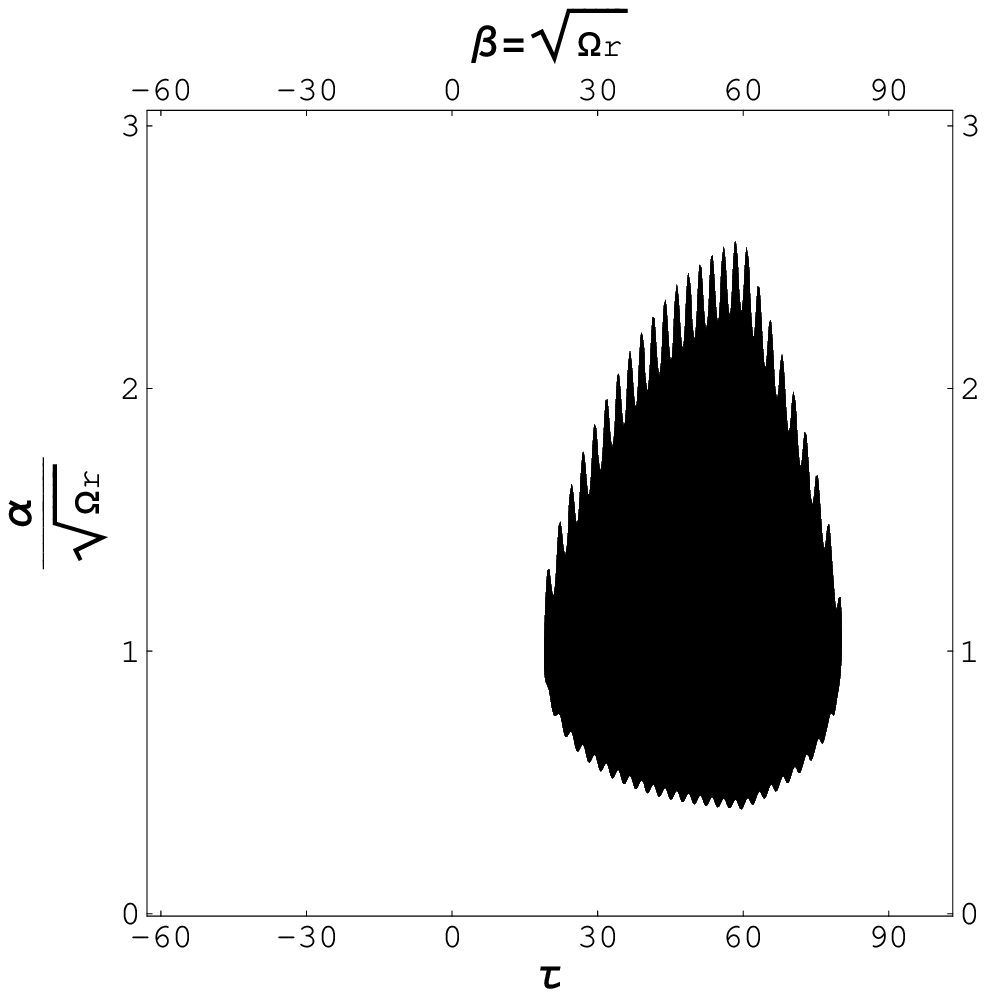}
\includegraphics[width=4cm]{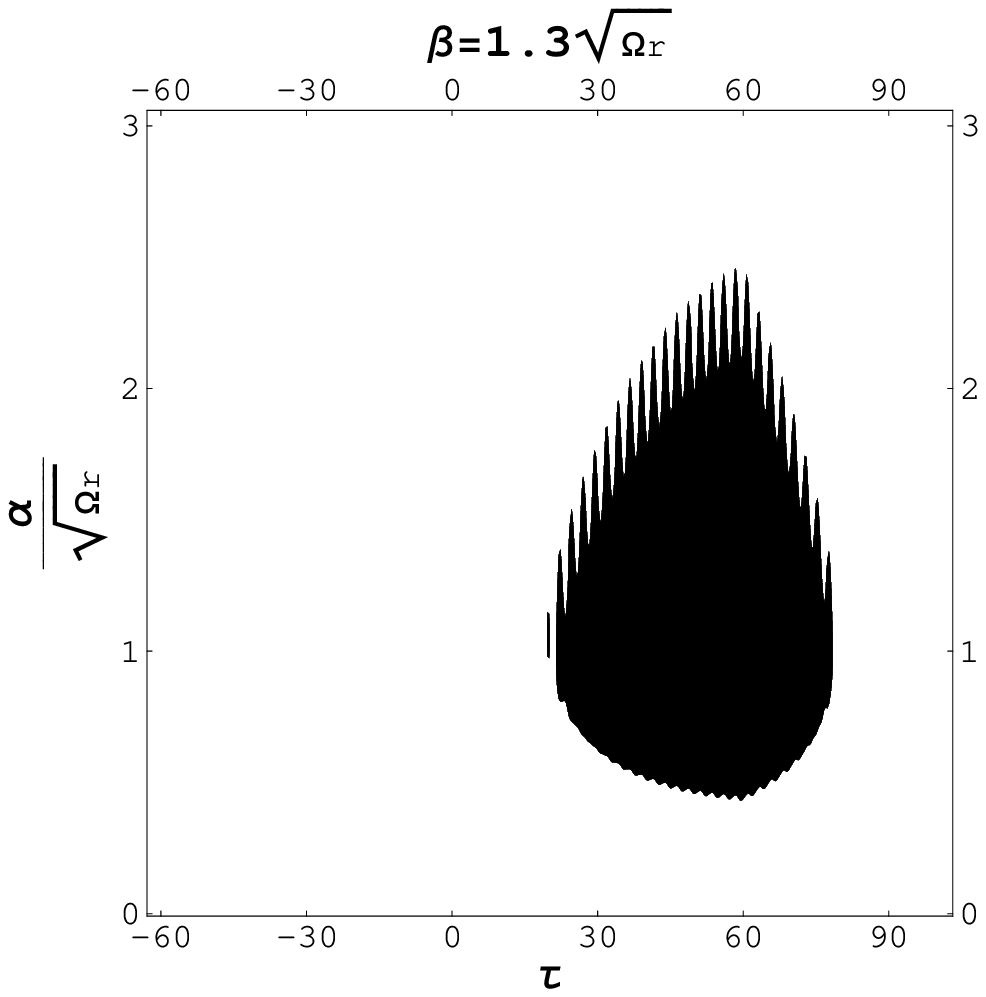}\\
\includegraphics[width=4cm]{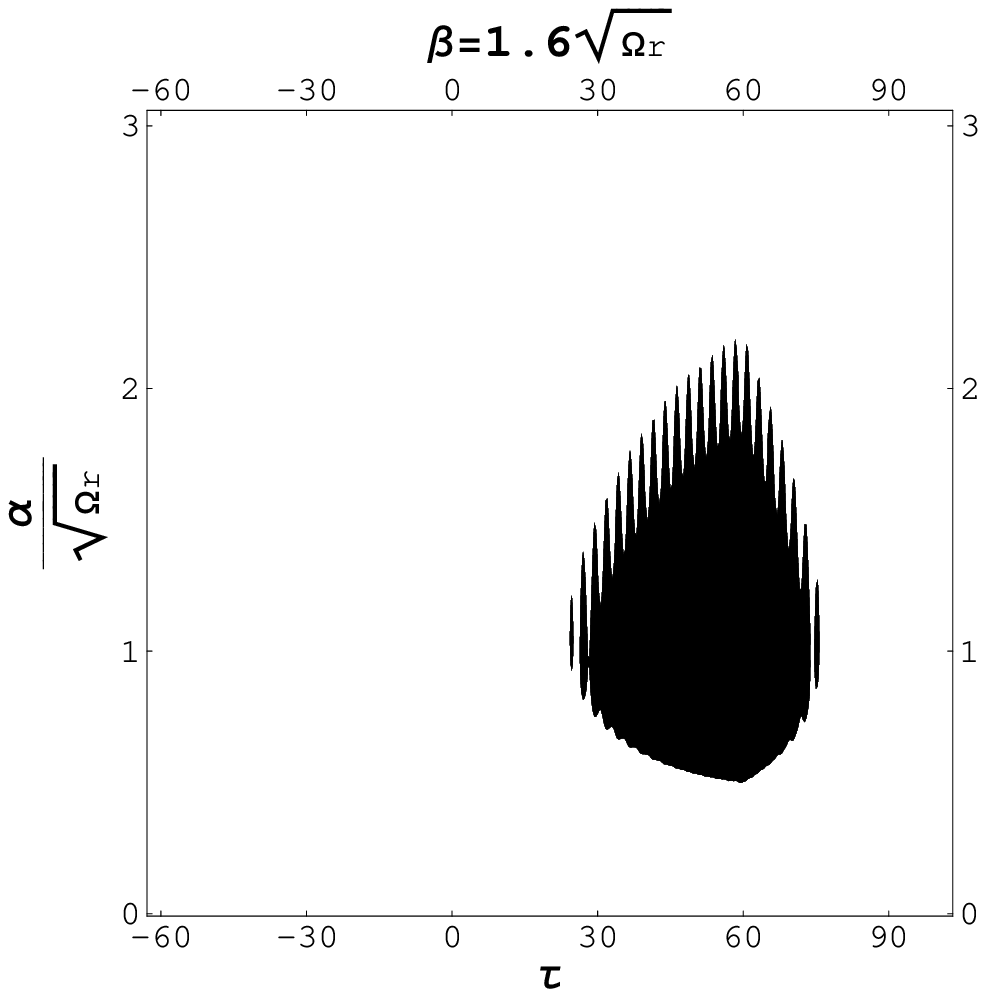}
\includegraphics[width=4cm]{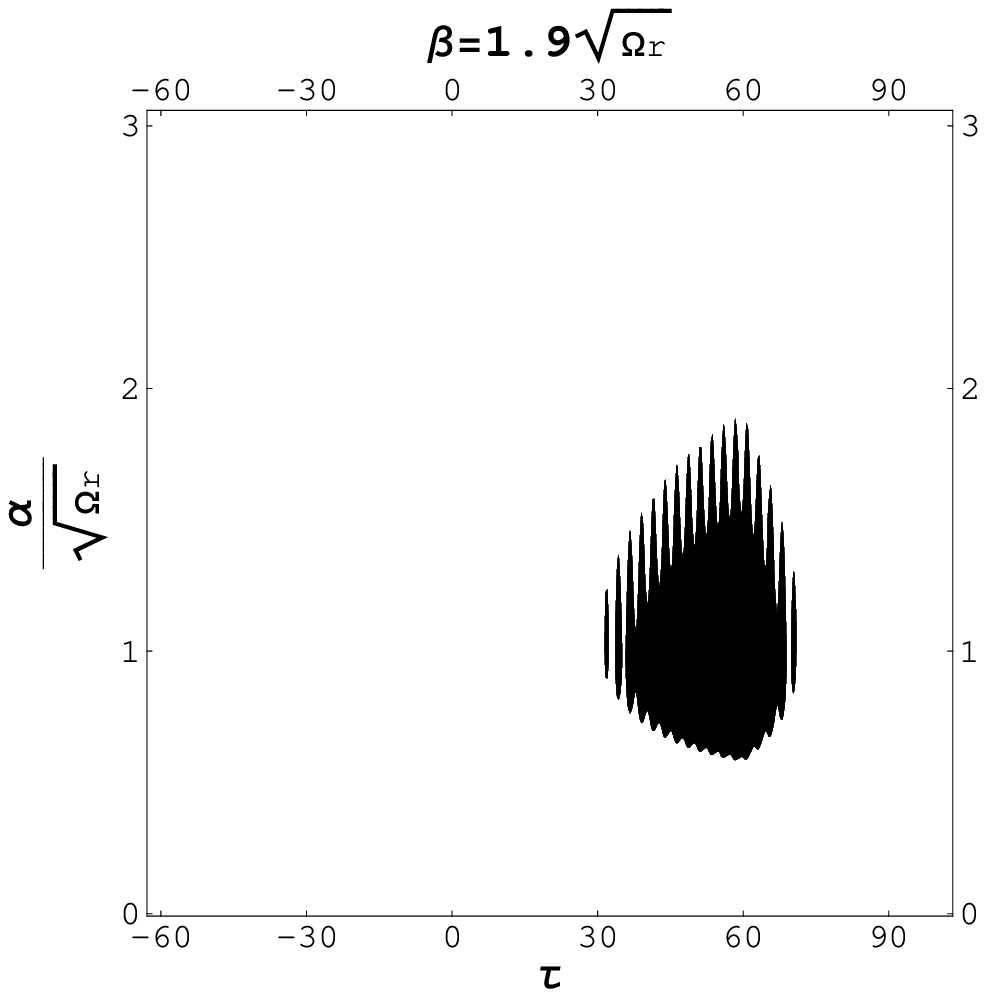}
\includegraphics[width=4cm]{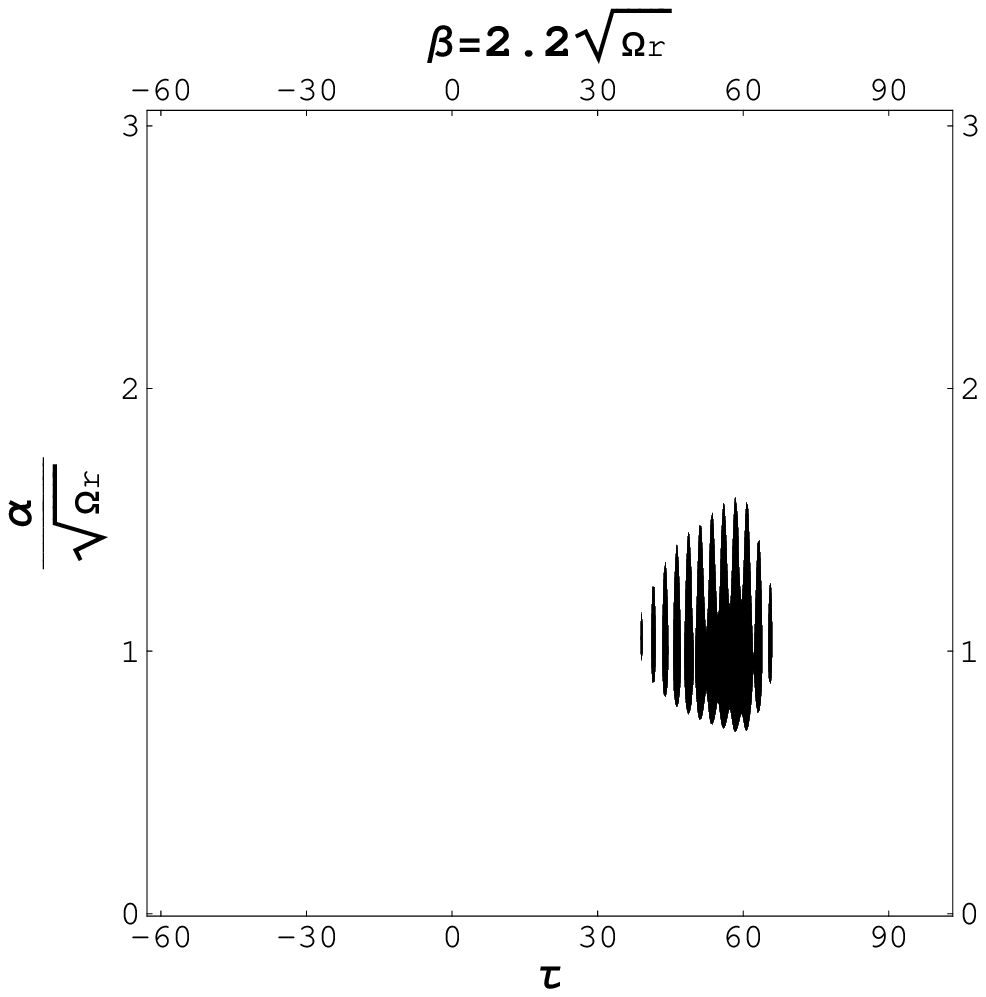}
\includegraphics[width=4cm]{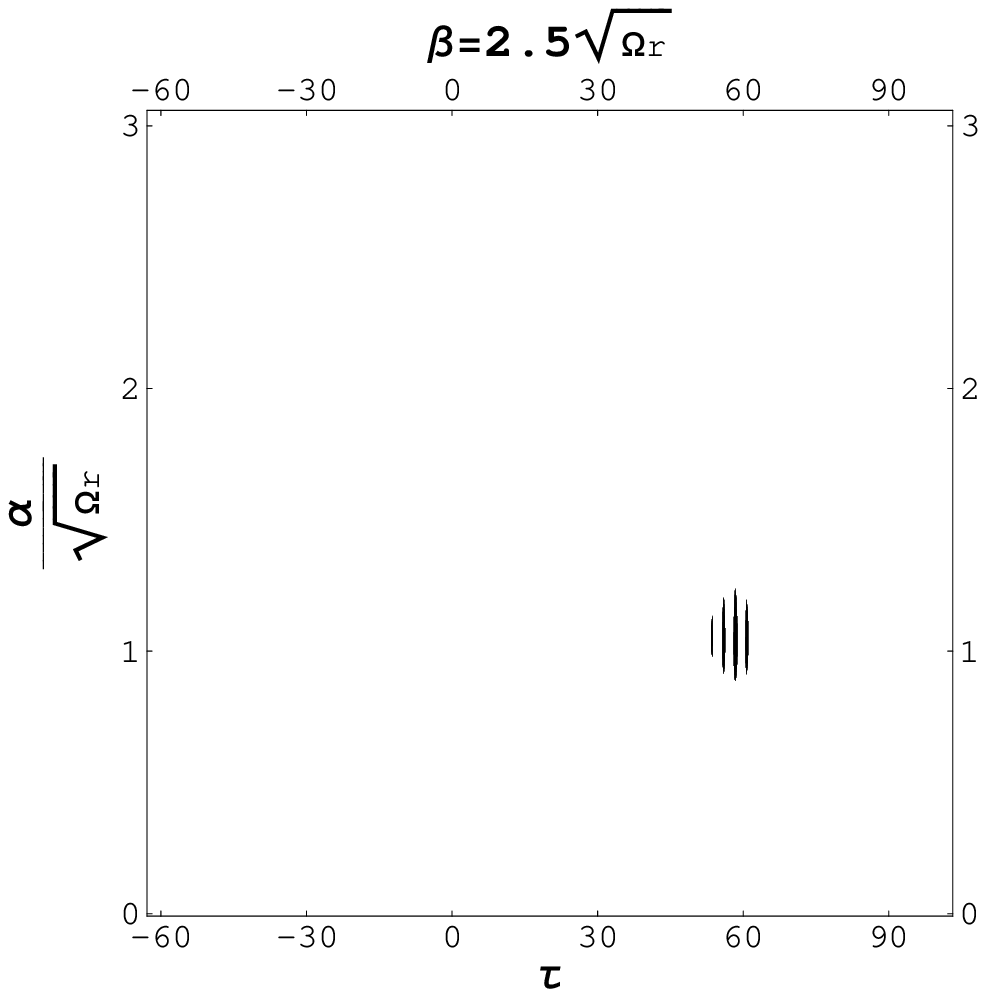}
\caption{Evolution of $\Sigma$ with each detector initially in a
squeezed state with minimum uncertainty. The parameters are the same as
those in Fig. \ref{QAQBdemo} (left) except $\alpha$ and $\beta$, which
are indicated in the plots. In the dark spots of the above plots,
$\Sigma < 0$ and the detectors are entangled. One can see that the
parameters in $(\alpha, \beta)$ space far away from the ones for the
ground state (at $(\alpha, \beta)=(\sqrt{\Omega_r},\sqrt{\Omega_r})$,
in the third plot from the left in the upper row)
tend to increase the value of $\Sigma$, namely, decrease the degree
of entanglement.} \label{Dyn001AlBe}
\end{figure}

For the cases with the initial state as a direct product of two
squeezed states of free detectors with minimum uncertainty, namely,
$(\alpha, \beta)\not= (\sqrt{\Omega_r},\sqrt{\Omega_r})$, the
only difference is the a-part of the self correlators. Detectors in
such cases will still be separable at late times, because all of the
a-part of the correlators eventually decays away, and the late-time
steady state is independent of the initial state of the detectors. In
transient, the a-part of the self correlators with the initial
squeezed state oscillates in time about the one with the initial
ground state: at some moment the former is greater than the latter,
and at another moment, lesser. But the oscillations of $\left<
\right.Q^2_j\left.\right>_{\rm a}$ and $\left<\right.P^2_j\left.
\right>_{\rm a}$ are out of phase, so the overall effect of the
a-part of the correlators is to increase the domination of the self
correlators and decrease the degree of entanglement (associated with
the value of $\Sigma$ or $E_{\cal N}$) if the values of $(\alpha,
\beta)$ are sufficiently far from $(\sqrt{\Omega_r}, \sqrt{\Omega_r})$
(or the squeeze parameters are sufficiently large; See Fig.
\ref{Dyn001AlBe}). From our numerical results we observe, however, that
it is still possible to enhance the tendency to entangle 
if one takes the values of $(\alpha, \beta)$ very close to the ones
for the initial ground state, while the enhancement is very tiny and
the correction to the values of $\Sigma$ is usually of the next order
in $\gamma$.

\subsection{Dynamics with Truncated RDM}

At early times in the perturbative regime the leakage of amplitudes
to higher excited states are negligibly small, so we expect the two
detectors can be approximately seen as a two 2-level systems (2LS).
For this kind of $2\times 2$ system the two 2LS are entangled if the
partial transpose of the RDM $\rho^R_{n_A n_B, n'_A n'_B}$ in
$(\ref{truncRho})$ has a negative eigenvalue, namely if one of the
inequalities
\begin{equation}
  \rho^R_{10,10}\rho^R_{01,01} - \rho^R_{11,00}\rho^R_{00,11} \ge 0,
  \label{keyineq}
\end{equation}
and
\begin{equation}
  \rho^R_{00,00}\rho^R_{11,11} - \rho^R_{10,01}\rho^R_{01,10} \ge 0,
\end{equation}
is violated. Reznik discovered \cite{Reznik03} that the inequality
$(\ref{keyineq})$ for the RDM of the two 2LS 
will always be violated in TDPT with
integration domain extended to ${\bf R}^2$
(denoted as TDPT$_\infty$; Cf. $(\ref{R1010TDPT})$-$(\ref{R1111TDPT})$), 
according to which he claimed that quantum entanglement can be
generated outside the light cone.

In Appendix \ref{RDM2dec} we derive the truncated RDM for two UD
detectors explicitly. Comparing this result with the full dynamics,
we can make the following observations:

1) Reznik's conclusion based on TDPT$_\infty$ is not generic for all
values of the initial moment $\tau_0$, the proper acceleration $a$,
the natural frequency $\Omega$, or the coupling strength $\gamma$.
Actually, the key result Eq.(18) in \cite{Reznik03} is correct only
at the moment $\tau=-\tau_0$ in the TDPT regime that entanglement
creation can occur. 

2) Beyond the TDPT$_\infty$ regime, our result shows that once
quantum entanglement is created, it only survives in a finite
duration. The two detectors are always separable at $\tau <0$ and at
late times.

3) The inequality $(\ref{keyineq})$ implies that, if the quantum state
of the detectors is separable, then from $(\ref{truncRho})$ one has
\begin{equation}
   0 \le {\cal J}^{AA'}{\cal J}^{BB'} - {\cal J}^{AB}{\cal J}^{A'B'}
   = {\Sigma \Omega^2 \det(V+V_0)\over 16 \hbar^{10} \left[
   \left<\right.Q_A^2\left.\right>\left<\right.Q_B^2\left.\right>
   -\left<\right.Q_A, Q_B\left.\right>^2\right]^2},
\end{equation}
which is a product of the quantity $\Sigma$ and a positive definite
function of time.
(Here $V_0\equiv (\hbar/2) {\rm diag}
(\Omega^{-1},\Omega, \Omega^{-1}, \Omega)$ is the covariance
matrix for two free detectors in their ground states.) Thus the
criterion of separability $(\ref{keyineq})$, though derived from the
RDM of the truncated oscillator, is actually equivalent to $\Sigma \ge 
0$, which is the sufficient and necessary criterion for Gaussian
states in the full dynamics \cite{LH2009}. In other words, the RDM of
the two oscillators each with a truncated spectrum, including only up
to the first excited state, possesses the complete information
pertaining to the separability of the two detectors in Gaussian states.

\begin{figure}
\includegraphics[width=8cm]{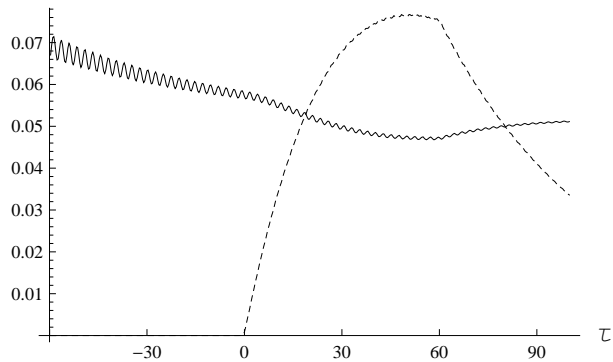}
\caption{Evolution of $|\rho^R_{11,00}|$ (dotted curve) and
$|\rho^R_{10,10}|$ (solid curve) of the truncated RDM under the same
condition as in Fig. \ref{exactDyn001} (left). The duration the
detectors are entangled determined by the condition
$|\rho^R_{11,00}|>|\rho^R_{10,10}|$ matches exactly with the result
determined by $c_- < \hbar/2$.}
\label{truncDyn001}
\end{figure}

We illustrate the time evolution of the exact $|\rho^R_{11,00}|$ and
$|\rho^R_{10,10}|$ from $(\ref{truncRho})$ in Fig. \ref{truncDyn001},
where the value of $|\rho^R_{11,00}|$ is highly dependent on the behavior
of the cross correlators (see Eqs. $(\ref{r1100WCL})$-$(\ref{aBmWCL})$):
it grows as the amplitude of the cross correlators grows. At the same
time the growing cross correlators tend to decrease the value of
$|\rho^R_{10,10}|$. Combining these two factors we find, in exactly
the same duration that the detectors are entangled in the full
dynamics, $|\rho^R_{11,00}|$ exceeds $|\rho^R_{10,10}|$ and violates
the inequality $(\ref{keyineq})$.

\section{Discussion}

\subsection{Summary of our findings}

We have studied the entanglement creation process of two uniformly
accelerated UD detectors moving in opposite directions in the
Minkowski vacuum of a massless scalar field. The two detectors are
causally disconnected during the whole history and are far apart at
late times.

For two detectors initially in their ground states, entanglement
creation does occur under some specific conditions: if the initial
time is negative and not very close to zero, the coupling strength
$\gamma$ between each detector and the field is not too large, and
the ratio of the proper acceleration $a$ to the natural frequency of
the detectors $\Omega$ is at some moderate value. The moment of 
entanglement creation, if any, is always at positive $\tau$.
Once quantum entanglement is created it can last for a lifetime much
longer than the natural period of the detectors in some parameter
range, while entanglement always disappears in a finite time.

For two detectors each initially in a squeezed state, similar
entanglement creation can also occur if the squeeze parameters in
the initial state of both detectors are sufficiently small so the
initial state is close enough to the direct product of the ground
states.

Moreover, we find that the RDM of two oscillators each with a truncated 
spectrum up to the first excited state contains the complete 
information about the separability of the oscillators in Gaussian states.
In Appendices A and B we see that in TDPT$_\infty$ regime, where the
integration domain has been extended to ${\bf R}^2$, the growing rates
of the matrix elements of the truncated RDM can be estimated well, but one
has to be careful in obtaining the ratio of different matrix elements.
While TDPT$_\infty$ results are Markovian, we find TDPT still
keeps some of the non-Markovian features.

                                                                                    %
The dependence of entanglement dynamics on the initial state, the fiducial 
time, and the non-Markovian features shown above are beyond the scope of 
Massar and Spindel's steady-state calculation \cite{MS06}. In addition, 
there are some other interesting differences between our findings in (3+1)D 
UD detector theory and MS's in (1+1)D U-RSG model. First, in our analysis 
$\Lambda_0$ and $\Lambda_1$ corresponding to the ultraviolet cutoffs in 
(3+1)D UD detector theory explicitly enter the entanglement dynamics, while 
the infrared cutoff in (1+1)D U-RSG model does not affect the degree of 
entanglement in MS's result because they only considered the cases where 
their constant $K$, which is a product of the coupling strength and the 
very long duration of the interaction corresponding to the infrared divergence, 
is always much greater than all other parameters, such that the higher-order 
terms in the $K$-expansion of the logarithmic negativity can be neglected.
Second, while in both cases quantum entanglement between the detectors can 
be created after the moment in Minkowski time that the distance between 
the detectors is a minimum, the lifetimes of the created entanglement in 
our case are usually much longer than the natural period of the detectors, 
in contrast to the short lifetimes of entanglement in MS's (1+1)D results. 
Finally, in our (3+1)D case entanglement creation is totally suppressed in 
the strong coupling regime and can manifest in the weak coupling limit, 
while in MS's result entanglement generation is a non-pertubative phenomena 
so there is no entanglement creation in their weak coupling regime. 
                                                                                    %

\subsection{Quantum nonlocality, causality, and reference frames}

Entanglement creation between two causally disconnected objects as we
have captured in this work may be viewed as \textit{a manifestation
of ``quantum nonlocality"} in quantum physics. Quantum entanglement
between two localized objects can be generated by allowing them to
interact with a common quantum field even though they do not exchange
any classical information. Foremost our results testify to the
important fact that \textit{quantum nonlocality does not violate
causality} \cite{CK09}.                                                           
Notice that quantum entanglement can be recognized and
put to use (as in QIP) only by those ``spectators" who can access the 
information from both detectors. From the viewpoint of the separate
detectors each can never find out the existence of the other from its
own RDM, nor the quantum entanglement between them. There is no 
transmittal of physical information between them and causality (of
information) is always respected.

This fact also means that the existence of a``spectator" is essential 
when we refer to the dynamics of entanglement between the two localized
parties outside of causal contact. The spectator is causally
connected to (and thus can ``see") both parties and can recognize the 
quantum entanglement between them through its own observation of both. 
Recall that in Ref. \cite{LCH2008} we found that entanglement
dynamics for two quantum objects in relativistic motion depend on the
choice of reference frames or coordinates (Minkowski or Rindler, for
example). One may debate on which coordinate is ``better" or more
``objective" for the depiction of entanglement dynamics. Our results
show that the only physically meaningful one in a given physical setup 
for describing the entanglement dynamics of the system is that of the 
spectator. This is of course not unique, but there are well-established 
ways to relate what is measured by one spectator to another with 
considerations of time-slicing (see, e.g., \cite{LCH2008}) and
transformation laws of reference frames in relativity theory.

\subsection{Quantum entanglement generated by collision}

The setup of our problem can also mimic the situation of two ions
of the same charge in a head-on collision. Quantum entanglement is 
generated in the later stage of the collision when the two ions are 
moving apart, once the trajectories of the ions possess the right 
kind of symmetry to enhance the field correlations as those 
in the present problem (see Appendix \ref{DpRidge}). 
There is no energy exchange between the two ions during the 
entanglement creation process of this kind. 

It has been shown that two atoms coupled with a common field vacuum 
in a cavity can get entangled in a ``collision" \cite{ZG00, Haro01}.
These entangling processes could be interpreted as a consequence of
virtual quanta exchange in a van der Waal potential, analogous to
those with the coulomb interaction in electron-electron scattering.
In \cite{Haro01} the interaction time is much longer than the
propagation time of photons across the spatial separation between the
atoms, and energy exchange between two atoms is clearly present.
These cases where entanglement generation occurring well inside the
light cone is different from the present setup but closer to the
situation we considered in \cite{LH2009}, where quantum entanglement
is established mainly by retarded mutual influences.

Note that particle (or atom, molecule) collisions in 
non-relativistic classical and quantum mechanics are usually described 
by an effective Hamiltonian with a direct, nonlocal, inter-particle
interaction (e.g. a potential $V(|{\bf r}_1 - {\bf r}_2|)$, where
${\bf r}_{1,2}$ are the positions of the particles). Direct
interaction always generates correlation between the particles,
classical or quantum. The nonlocality of the interaction, on the
other hand, is a consequence of coarse graining from a more
fundamental local theory, resulting in an effective theory 
description.  The particles in this non-relativistic regime cannot
resolve the time scale for information propagating back and forth
between them. Thus the generation of correlation by this kind of
nonlocal direct interaction happens in a time scale during which two
particles have causal contact for a long time. One cannot tell
whether correlations can be created between two causally disconnected
particles in this kind of effective theories. Moreover, with this
kind of nonlocal direct interaction, if one does not further
introduce a spatial range of interaction associated with a 
coarse-graining in time, even classical correlation will be nonlocal 
and cannot be described by any local hidden-variable theory.
\\

\noindent{\bf Acknowledgment} SYL wishes to thank Daniel Braun
for illuminating discussions during his visit to
the Joint Quantum Institute, University of Maryland and National
Institute of Standards and Technology, where this work was first
motivated and commenced. BLH wishes to thank the hospitality of National 
Center for Theoretical Sciences and the QIS group at National Cheng Kung 
University of Taiwan. This work is supported partially by grants from 
LPS, NSF grants PHY-0426696, PHY-0801368 and DARPA-HR0011-09-1-0008.

\begin{appendix}

\section{Matrix elements in time-dependent perturbation theory}
\label{TDPTapx}

In TDPT, one expands the wave function of the two detectors in terms of the
energy eigenstates $\varphi_{m}(Q_A)$ and $\varphi_{n}(Q_B)$ of the free
detectors as
\begin{equation}
 \psi(\tau) = \sum_{m', n'} C_{m'n',mn}e^{-i(\omega_{m'}+\omega_{n'})\tau
  -i(\omega_m+\omega_n)\tau_0}\varphi_{m'}(Q_A)\varphi_{n'}(Q_B),
\end{equation}
then calculates the first few factors in the perturbative series of
$C_{m'n',mn}$ assuming a small coupling constant $\lambda_0$, namely,
$C_{m'n',mn} \equiv \sum_N \lambda_0^N C_{m'n',mn}^{(N)}$. Here
$\omega_n \equiv E_n/\hbar = \Omega[n+ (1/2)]$. For the initial state
$(\ref{initGauss})$, to lowest order in $\lambda_0$, the elements of
the RDM of the detectors are
\begin{eqnarray}
  \rho^R_{10,10} &\approx& {\lambda_0^2 \over 2 \hbar \Omega}
    \int_{\tau_0}^\tau ds\int_{\tau_0}^{\tau} ds'
    e^{-i\Omega(s-s')} D^+\left(z^\mu_A(s), z^\nu_A(s')\right),
    \label{R1010TDPT}\\
  \rho^R_{10,01} &\approx& {\lambda_0^2 \over 2 \hbar \Omega}
    \int_{\tau_0}^\tau ds\int_{\tau_0}^{\tau} ds'
    e^{-i\Omega(s-s')} D^+\left(z^\mu_A(s), z^\nu_B(s')\right),
    \label{R1001TDPT}\\
  \rho^R_{11,00} &\approx& -{\lambda_0^2 e^{-2i\Omega \tau}\over
    2\hbar\Omega} \int_{\tau_0}^\tau ds\int_{\tau_0}^{\tau} ds'
    e^{i\Omega(s+s')} D^+\left(
    z^\mu_A(s), z^\nu_B(s')\right), \label{R1100TDPT}\\
  \rho^R_{11,11} &\approx&  \left({\lambda_0^2 \over 2\hbar\Omega}\right)^2
    \int_{\tau_0}^\tau ds ds' d\bar{s}d\bar{s}'
    e^{i\Omega(s+s'-\bar{s}-\bar{s}')}
    \left[ D^+\left(z^\rho_A(\bar{s}), z^\sigma_B(\bar{s}')\right)
    D^+\left(z^\mu_A(s), z^\nu_B(s')\right)+ \right.\nonumber\\ & & \left.
    D^+\left(z^\rho_A(\bar{s}), z^\mu_A(s)\right)
    D^+\left(z^\sigma_B(\bar{s}'), z^\nu_B(s') \right)+
    D^+\left(z^\rho_A(\bar{s}), z^\nu_B(s')\right)
    D^+\left(z^\sigma_B(\bar{s}'), z^\mu_A(s) \right)\right],
  \label{R1111TDPT}
\end{eqnarray}
etc., where $D^+$ is the positive frequency Wightman function of the
massless scalar field in Minkowski vacuum, given by
\begin{eqnarray}
  D^+(z_j^\mu,z_{j'}^\nu) &\equiv&
    \left<\right. 0_M\left.\right|\phi\left(z^\mu_j\right)
    \phi\left(z^\nu_{j'}\right) \left| \right. 0_M \left.\right> \nonumber \\
  &=& \int {\hbar d^3 k\over (2\pi)^3 2\omega}e^{-\omega\epsilon}
    e^{-i \omega (z_j^0 - z_{j'}^0)+ i {\bf k}\cdot({\bf z}_j - {\bf z}_{j'})}
  \nonumber\\ &=& {\hbar/4\pi^2\over\left|{\bf z}_j-{\bf z}_{j'}\right|^2-
  \left(z_j^0 -z_{j'}^0-i\epsilon\right)^2}
\label{wightgen}
\end{eqnarray}
with $j,j'=A,B$ and the frequency $\omega \equiv|{\bf k}|$ for the massless
scalar field. $\epsilon$ is a small positive real number serving as a
regulator at high frequency in ${\bf k}$-integration.

\subsection{The regulators}
\label{regu}

In obtaining $\rho^R_{1010}$ in $(\ref{R1010TDPT})$, one could insert
the trajectory of the detector A into $(\ref{wightgen})$, which gives the
Wightman function as
\begin{equation}
  D^+(z_A^\mu(s),z_A^\nu(s')) = 
    {\hbar/4\pi^2\over -{4\over a^2}\sinh {a\over 2}\Delta\left(
    \sinh {a\over 2}\Delta- i\epsilon a\cosh aT \right)+ \epsilon^2},
\label{originalDp}
\end{equation}
with $T\equiv(s+s')/2$ and $\Delta\equiv s -s'$.
Conventional wisdom says that the value of $\epsilon$ is extremely
small, so the expression of the Wightman function could be replaced by
\cite{BD82}
\begin{eqnarray}
  & & D'^+(z_A^\mu(s),z_A^\nu(s')) = {\hbar/4\pi^2\over
    -{4\over a^2}\sinh^2 {a\over 2}(\Delta- i\epsilon')} \nonumber\\
    &\approx& {\hbar/4\pi^2\over
    -{4\over a^2}\sinh {a\over 2}\Delta \left( \sinh {a\over 2}\Delta
      - i\epsilon' a \cosh {a\over 2}\Delta\right)+ \epsilon'^2\cosh a\Delta},
\label{modifiedDp}
\end{eqnarray}
which is independent of $T$. Indeed, when $aT$ is small, the original $D^+$
and the modified $D'^+$ have similar positions of poles and similar
behavior around the poles in the complex $\Delta$ plane.

\begin{figure}
\includegraphics[width=8cm]{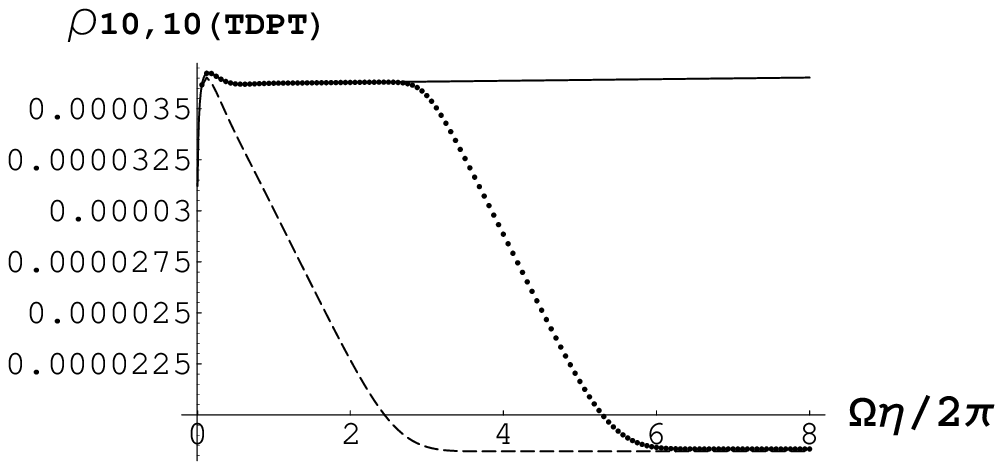}
\includegraphics[width=8cm]{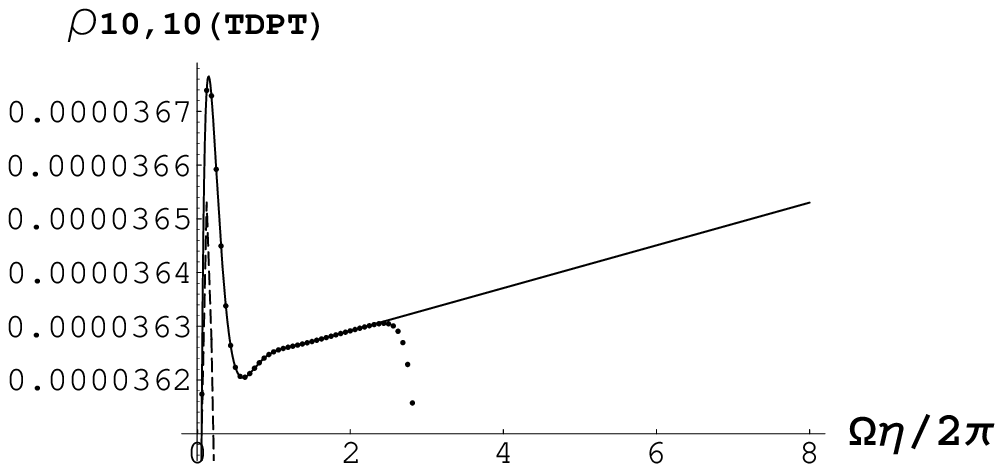}
\caption{Early-time evolutions of the $\rho_{10,10}^R$ in TDPT with
the original positive frequency Wightman function,
$D^+(z_A^\mu(s),z_A^\nu(s'))$ in Eq. $(\ref{originalDp})$ (dashed
curve) with $\epsilon = e^{-14 - \gamma_e}/\Omega$ ($\gamma_e$ is the
Euler's constant), the $\rho_{10,10}^R$ with the modified one,
$D'^+(z_A^\mu(s), z_A^\nu(s'))$ in Eq. $(\ref{modifiedDp})$ (solid
curve) with $\epsilon' = e^{-14 - \gamma_e}/\Omega$, and the
$\rho_{10,10}^R$ in $(\ref{R1010phys})$ with the original
$D^+(z_A^\mu(s),z_A^\nu(s'))$ and $\epsilon= e^{-30}/\Omega$,
together with the physical cutoffs $\epsilon_0=\epsilon_1= e^{-14 -
\gamma_e}/\Omega \gg \epsilon$ (dots). The left and the right plots
are the same results but shown in different scales to stress that the
early-time behavior of the dotted curve  really matches with the
climbing solid curve. Here $\gamma=10^{-5}$ and $\Omega=2.3$. The
slope of the dotted curve in its linear growing phase agrees quite
well with $2\gamma/(e^{2\pi\Omega/a}-1)$ in $(\ref{R1010PTinf})$.}
\label{compareDp}
\end{figure}

However, as shown in Fig. \ref{compareDp}, only the modified $D'^+$ will
give a linear-growing phase in transient like
\begin{equation}
  \rho_{10,10}^{R\infty} = 2\gamma\eta/(e^{2\pi\Omega/a}-1),
\label{R1010PTinf}
\end{equation}
where the growing rate $\rho_{10,10}^{R\infty}/\eta$ can be obtained by
extending the integration domain of $(s, s')$ to ${\bf R}^2$, namely,
from TDPT$_\infty$. In contrast, the original $D^+$
leads to a totally different behavior of $\rho_{10,10}^R$: It is
linearly decreasing in the beginning and saturates after $\tau >
-a^{-1}\ln \epsilon$.

Learning from what we did in taking the coincidence limit of the self
correlators \cite{LH2006}, we find that, if we put the small shifts
$\epsilon_0$ and $\epsilon_1$ at the upper and lower bounds of the
$s$ and $s'$ integration domain, such as
\begin{equation}
  \rho^R_{10,10} \approx {\lambda_0^2 \over 2 \hbar \Omega}
    \int_{\tau_0}^{\tau+\epsilon_1} ds\int_{\tau_0+\epsilon_0}^{\tau} ds'
    e^{-i\Omega(s-s')} D^+\left(z^\mu_A(s), z^\nu_A(s')\right),
    \label{R1010phys}
\end{equation}
and take the $\epsilon$ in $D^+$ to zero limit, then the exotic
behavior of $D^+$ will be altered and we recover the results obtained
by using the modified $D'^+$ with $\epsilon'$ corresponding to
$\epsilon_0$ when $\epsilon_0=\epsilon_1$. So we interpret $\epsilon$
as the mathematical cutoff, which should be taken the zero limit at
some point of calculation, while the $\epsilon_0$, $\epsilon_1$ in
$(\ref{R1010phys})$ and the $\epsilon'$ in $(\ref{modifiedDp})$ are
physical cutoffs corresponding to the time resolution of the
detectors and related to the constants $\Lambda_0$ and $\Lambda_1$ in
the expressions for the self correlators \cite{LH2006}.
The replacement from $(\ref{originalDp})$ to $(\ref{modifiedDp})$
actually changes the interpretation of the regulator.

\subsection{Calculating $|\rho^R_{11,00}|$}
\label{DpRidge}

Usually in obtaining $\rho^R_{10,10}$, TDPT is good from
$\eta\equiv\tau-\tau_0 \gg \Lambda_1/\Omega$ up to $\eta\sim
O(1/\gamma)$ \cite{LH2006} with $\gamma\equiv\lambda_0^2/8\pi$ (here
the masses of both harmonic oscillators in the detectors have been
set to $m_0=1$ and the proper accelerations of the detectors are not
extremely large). This is a large duration if $\gamma$ is small,
which is consistent with the assumption of TDPT, so it is commonly
argued that extending the domain of $s$ integration from $(\tau_0,
\tau)$ to $( -\infty, \infty)$ in this duration would not change the
results too much, that is, the TDPT$_\infty$ results should be very
close to the TDPT results.

In Ref.\cite{Reznik03}, Reznik compared the absolute values of
$|\rho^R_{11,00}|$ and $|\rho^R_{10,10}|$ in TDPT$_\infty$ and
discovered that the former is always greater than the latter, then
drew the conclusion that the two detectors are always entangled.
Nevertheless, with the extended integration domain, both
$|\rho^R_{11,00}|$ and $|\rho^R_{10,10}|$ are infinities, which are
meaningless as elements of any density matrix.
Usually one divides $|\rho^R_{10,10}|$ by the infinite duration of
interaction and interprets the result as the transition probability per
unit time in the linear-growing phase in TDPT regime \cite{BD82}.
But the choice of proper duration for $|\rho^R_{11,00}|$ and the
interpretation for the result could be tricky
(see Appendix \ref{RDM2dec} for more details).
Further, our numerical results show that if we calculate $|\rho^R_{11,
00}|$ with a finite integration domain and a nonzero regulator
$\epsilon$, the value of $(\ref{R1100TDPT})$ after integration can be
quite different from those obtained in Ref.\cite{Reznik03}.

\begin{figure}
\includegraphics[width=8cm]{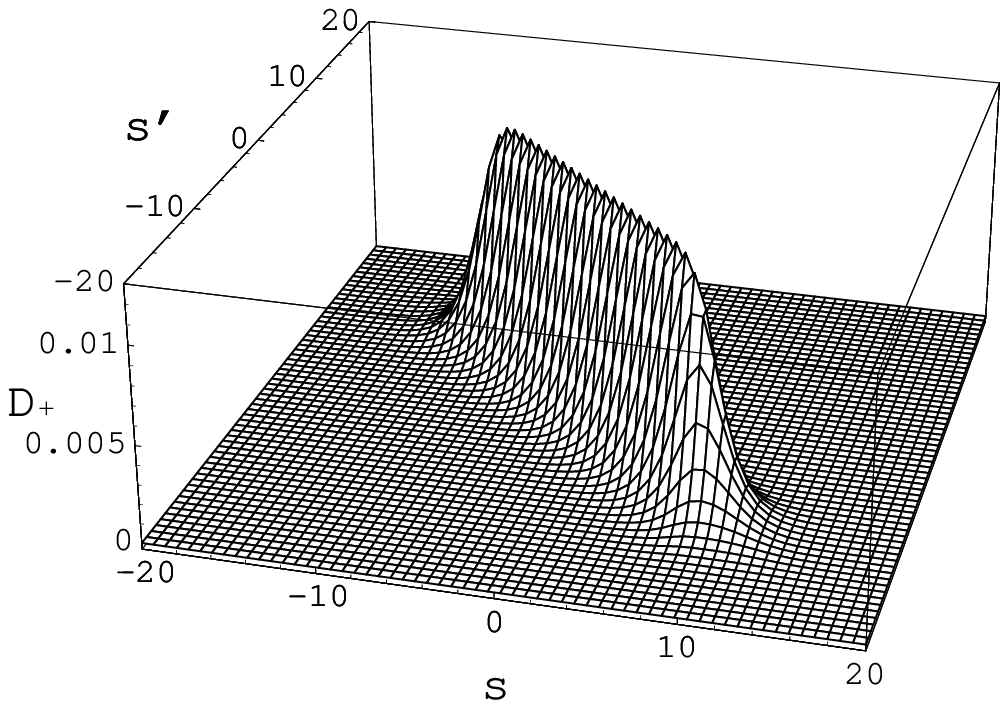}
\includegraphics[width=7cm]{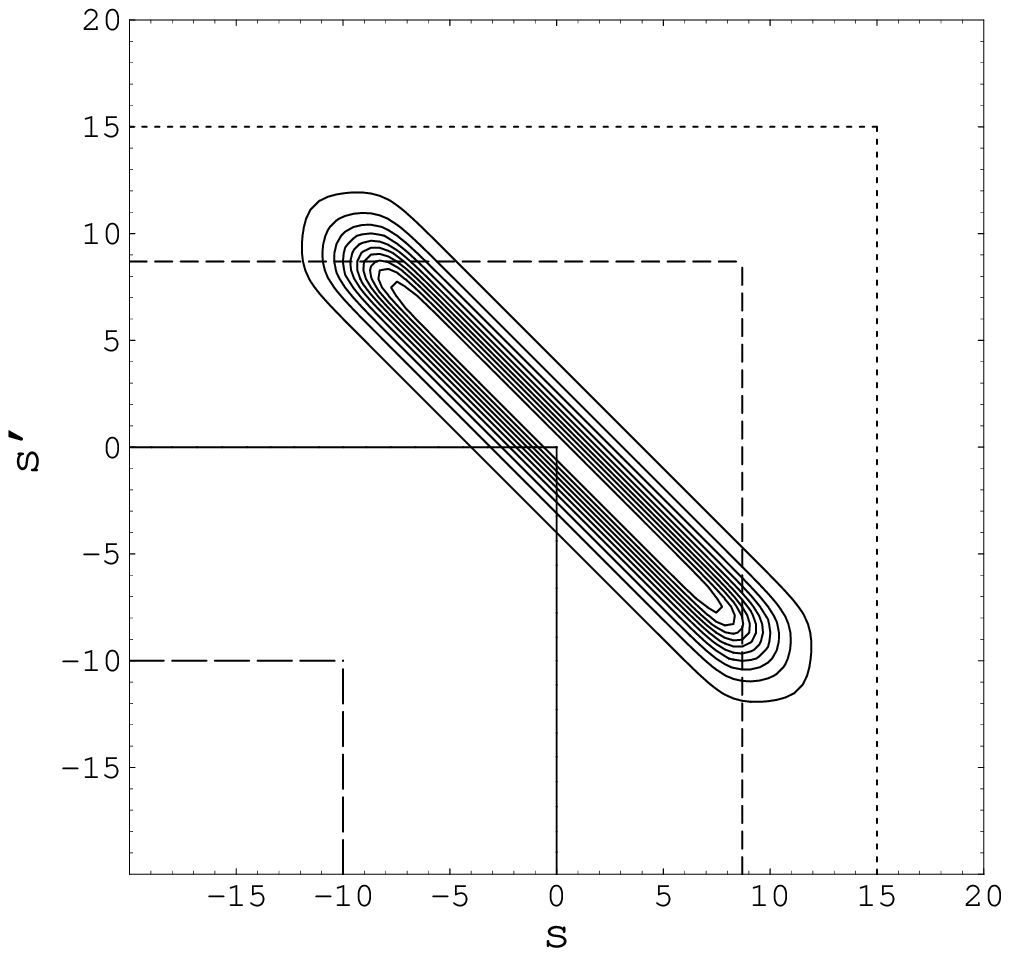}
\caption{(Left) An example of $|D^+(z_A^\mu(s),z_B^\nu(s'))|$ in Eq. 
$(\ref{DpAB})$. Here $\epsilon = e^{-8}$ and $a=1$. One can see that
the correlation of the massless scalar field at $z_A^\mu(s)$ and 
$z_B^\nu(s')$ is enhanced around $s'=-s$. Such an enhancement will be 
suppressed if the trajectories of the detectors are off the $z_A^\mu
(s)$ and $z_B^\nu(s')$ in this paper or the scalar field is massive. 
(Right) Contour plot of the same $|D^+(z_A^\mu(s),z_B^\nu(s'))|$.
The long-dashed lines, solid lines, short-dashed lines, and dotted lines
are borders of the integration domain at $\tau=-10$, $0$, $1/a \sinh^{-1}
\epsilon^{-1}(\approx 8.69)$, and $15$, respectively.}
\label{wightman}
\end{figure}

To calculate $\rho^R_{11,00}$, one substitutes the trajectories
$z_A^\mu$ and $z_B^\mu$ into the positive frequency Wightman function
and obtain
\begin{equation}
  D^+(z_A^\mu(s),z_B^\nu(s')) =
  {\hbar /4\pi^2\over  {4\over a^2}\cosh aT \left( \cosh aT +i \epsilon a
  \sinh {a\over 2}\Delta \right)+\epsilon^2}. \label{DpAB}
\end{equation}
Now $D^+(z_A^\mu(s),z_B^\nu(s'))$ is regular everywhere in $(s,s')$
plane. For small finite $\epsilon$, since $\cosh aT \ge 1$, the 
$\epsilon^2$ term in the denominator can always be neglected, but the 
$\epsilon$ term cannot. If $\epsilon \sinh a\Delta /2 \ll \cosh aT$, 
then $D^+\approx a^2/16\pi^2 \cosh^2 aT$, which is a function of $T$ 
only and peaks around $T=0$. For $\Delta$ sufficiently large such that
$\epsilon a\sinh a\Delta/2\agt 1$, however, $|D^+|$ will be suppressed. 
So $|D^+(z_A^\mu(s),z_B^\nu(s'))|$ looks like a ridge at $T\approx 0$ 
and spread with the width $4\tau_1$ in $\Delta$ direction, where 
$\tau_1\equiv (1/a)\sinh^{-1} (1/a\epsilon) \approx (1/a) \ln 
(2/a\epsilon)$ (see Fig. \ref{wightman}).

Suppose the coupling is switched on at the initial moment $\tau_0<0$.
From the landscape of $D^+(z_A^\mu(s),z_B^\nu(s'))$ shown in Fig.
\ref{wightman}, the evolution of $|\rho^R_{11,00}|$ will have three stages:

(i) When $\tau_0< \tau < 0$, $D^+(z_A^\mu(s),z_B^\nu(s'))$ is small
all over the ${\bf s}$-integration domain (within the square with two
sides in long-dashed lines in Fig. \ref{wightman} (right)), so the
change in $|\rho^R_{11,00}|$ is tiny.

(ii) At about $\tau =0$, the edge of the ${\bf s}$-integration domain
[within the square with all sides in solid lines in Fig.
\ref{wightman} (right)] touches the ridge of $|D^+(z_A^\mu(s),
z_B^\nu(s'))|$ around $T=0$, which gives obvious contribution to the
result. As $\tau$ grows, more and more portion of the ridge are
included in the integration domain. Since the height of the ridge is
almost constant in the $\Delta$ direction, $|\rho^R_{11,00}|$ grows
linearly in $\tau$.

(iii) The linear growth terminates at about $\tau = \tau_1$. Then the
integration domain has covered almost all the ridge so that $|D^+(
z_A^\mu(s), z_B^\nu(s'))|$ does not add obvious contribution after $\tau =
\tau_1$ and $|\rho^R_{11,00}|$ saturates at some constant.

From the argument in Sec. \ref{regu}, one should take $\epsilon
\to 0$ so $\tau_1$ and the width of the ridge in the $\Delta$ direction
is virtually infinite. Then stage (ii) will be terminated around $\tau
\approx |\tau_0|$ because only a portion of the ridge with $|\Delta|
\alt 2|\tau_0|$ is covered in the integration domain. This could be
enough to make the value of $|\rho^R_{11,00}|$ exceeds
$|\rho^R_{10,10}|$ and the two detectors get entangled, if $|\tau_0|$
is sufficiently large. On the other hand, for $\tau_0 \ge 0$, the
linear growth of stage (ii) never occurs, and $|\rho^R_{11,00}|$ is
always less than $|\rho^R_{10,10}|$.

Thus one can see that the value of $|\rho^R_{11,00}|$ as well as the
ratio $|\rho^R_{11,00}|/|\rho^R_{10,10}|$ have complicated evolution in 
time even in TDPT. Their behaviors are quite different from the constant
ones obtained in Ref.\cite{Reznik03}.

\section{Truncated reduced density matrix of the two detectors}
\label{RDM2dec}

Generalize the method applied in \cite{LH2006} for one single detector,
the elements of the RDM in eigen-energy representation can be read off by
comparing the coefficients of the $s_j^{n_j}$ terms in both sides of the
following equation,
\begin{equation}
   \sum_{n_i =0}^\infty \left[\rho^R_{n_A n_{A'}, n_B n_{B'}}
     \prod_j s_j^{n_j}\sqrt{{2^{n_j}\over n_j!} } \right]
   = \sqrt{ \prod_j K_j\over {\cal G}\det \tilde{G}}
     \, \exp \sum_{i,j} s_i \left[ K_i (\tilde{G}^{-1})^{ij}K_j
     -\delta^{ij}\right]s_j ,
\end{equation}
where $i,j=A,A',B,B'$, $K_A=K_{A'}=K_B=K_{B'}=K\equiv \sqrt{\Omega_r /\hbar}$,
${\cal G}$ is defined by
\begin{equation}
   {\cal G}\equiv 4 \left[\left<\right.Q_A^2\left.\right>
   \left<\right.Q_B^2\left.\right>-\left<\right.Q_A,Q_B\left.\right>^2\right],
\label{calG}
\end{equation}
and $\tilde{G}$ is parametrized in
\begin{equation}
  \tilde{G} \equiv \left( \begin{array}{cccc}
    a_{A+}+i b_A+{K_A^2\over 2} & a_{A-} & a_{X+}+i b_{X+} & a_{X-}+i b_{X-} \\
    a_{A-} & a_{A+}-i b_A+{K_{A'}^2\over 2}& a_{X-}-i b_{X-} & a_{X+}-i b_{X+} \\
    a_{X+}+i b_{X+} & a_{X-}-i b_{X-} & a_{B+}+i b_B+{K_B^2\over 2}& a_{B-} \\
    a_{X-}+i b_{X-} & a_{X+}-i b_{X+} & a_{B-} & a_{B+}-i b_B+{K_{B'}^2\over 2}
    \end{array}\right),
\end{equation}
with the factors
\begin{eqnarray}
  a_{A\pm} &=& {1\over 2{\cal G}}\left[ V^{11} \pm
    {4\over \hbar^2}\left(V^{-1}\right)^{22}\det V\right],\\
  a_{B\pm} &=& {1\over 2{\cal G}}\left[ V^{33} \pm
    {4\over \hbar^2}\left(V^{-1}\right)^{44}\det V\right],\\
  a_{X\pm} &=& -{1\over 2{\cal G}}\left[ V^{13} \pm
    {4\over \hbar^2}\left(V^{-1}\right)^{24}\det V\right],\\
  b_A &=& {2\over\hbar{\cal G}}\left(
    \left<\right.P_A,Q_B\left.\right>\left<\right.Q_A,Q_B\left.\right>
   -\left<\right.P_A,Q_A\left.\right>\left<\right.Q_B^2\left.\right>\right),\\
  b_B &=& {2\over\hbar {\cal G}}\left(
    \left<\right.P_B,Q_A\left.\right>\left<\right.Q_A,Q_B\left.\right>
   -\left<\right.P_B,Q_B\left.\right>\left<\right.Q_A^2\left.\right>\right),\\
  b_{X\pm} &=& {1\over\hbar {\cal G}}\left[
   \left(\left<\right.P_A,Q_A\left.\right>\left<\right.Q_A,Q_B\left.\right>
   -\left<\right.P_A,Q_B\left.\right>\left<\right.Q_A^2\left.\right>\right)
   \pm \right. \nonumber\\ & & \,\,\, \left.
   \left(\left<\right.P_B,Q_B\left.\right>\left<\right.Q_A,Q_B\left.\right>
   -\left<\right.P_B,Q_A\left.\right>\left<\right.Q_B^2\left.\right>\right)
   \right].
\end{eqnarray}
These $a$ and $b$ factors are obtained by solving the coefficients $G^{ij}$
in the Gaussian RDM of the detectors $\rho^R \sim \exp -\sum_{i,j}Q_i
G^{ij}Q_j$ in terms of the two-point correlators.
Here $V^{ij}$ are the elements of the covariance matrix
\begin{equation}
  V = \left( \begin{array}{cc} {\bf v}_{AA} & {\bf v}_{AB} \\
                   {\bf v}_{BA} & {\bf v}_{BB} \end{array}\right),
\label{CovarMtx}
\end{equation}
in which the elements of the $2\times 2$ matrices ${\bf v}_{ij}$ are
symmetrized two-point correlators ${\bf v}_{mn}{}^{\mu\nu} =
\left<\right.{\cal R}^\mu_m , {\cal R}^\nu_n \left.\right> \equiv
\left<\right.({\cal R}^\mu_m {\cal R}^\nu_n + {\cal R}^\nu_n {\cal R}^\mu_m )
\left.\right>/2$ with ${\cal R}^\mu_m = (Q_m(t), P_m(t))$, $\mu,\nu= 1,2$
and $m, n = A, B$.

Up to the first excited states of both detectors, the truncated RDM of the
two detectors in eigen-energy representation reads
\begin{equation}
  \rho^R_{n_A n_B, n'_A n'_B} \approx\left( \begin{array}{cccc}
    g & 0 & 0 & gK^2 {\cal J}^{A'B'} \\
    0 & gK^2 {\cal J}^{BB'} & gK^2 {\cal J}^{A'B} & 0 \\
    0 & gK^2 {\cal J}^{AB'} & gK^2 {\cal J}^{AA'} & 0 \\
    gK^2 {\cal J}^{AB} & 0 & 0 & gK^4 \left[
    {\cal J}^{AB}{\cal J}^{A'B'}+ {\cal J}^{AA'}{\cal J}^{BB'}+
    {\cal J}^{AB'}{\cal J}^{A'B}\right]  \end{array}\right)
\label{truncRho}
\end{equation}
with $n_A, n_B, n'_A, n'_B = 0,1$, ${\cal J}\equiv \tilde{G}^{-1}$, and
\begin{equation}
   g \equiv {\Omega/\hbar \over \sqrt{{\cal G}\det \tilde{G}}}.
\end{equation}
Here we use the basis $(n_A,n_B)=\{00,01,10,11\}$ for $\rho^R$.

What TDPT$_\infty$ could describe well is the transient behavior during
$\Omega^{-1} \ll \tau < |\tau_0| \ll 1/2\gamma$ while $\Omega^{-1}\ll
\tau-\tau_0 \ll 1/\gamma$ with $\tau_0 <0$, when the amplitudes of the
oscillating cross correlators are linearly increasing, while the
magnitudes of the self correlators are still in transient and growing
linearly, too. In this regime, $\left<\right. Q_A,Q_B\left.\right>$
can be approximated by $(\ref{t0ggt})$, which also yields
\begin{eqnarray}
  & & \left<\right. P_A,Q_B\left.\right> \approx
  \left<\right. Q_A,P_B\left.\right> \approx
  {\hbar\gamma\over\Omega \sinh{\pi\Omega\over a}}\left[ 2\Omega\tau
  \sin 2\Omega\tau + \left({\pi\Omega\over a}\coth{\pi\Omega\over a}
  -1\right)\cos 2\Omega\tau\right], \\
  & & \left<\right. P_A,P_B\left.\right> \approx
  {\hbar\gamma\over \sinh{\pi\Omega\over a}}\left[ 2\Omega\tau
  \cos 2\Omega\tau - \left({\pi\Omega\over a}\coth{\pi\Omega\over a}
  -1\right)\sin 2\Omega\tau\right],
\end{eqnarray}
up to $O(\gamma)$. Together with the $O(\gamma)$ results of the self
correlators in TDPT regime from \cite{LH2006} with $m_0=1$,
\begin{eqnarray}
  \left<\right. Q_A^2\left.\right> =\left<\right. Q_B^2\left.\right>
  &\approx& {\hbar\over 2\Omega} +\nonumber\\
  & & {\hbar\gamma\over\Omega}\left[
  \left(\coth {\pi\Omega\over a}-1\right)\eta+
  {2\over\pi\Omega}(\Lambda_0-\ln a)\sin^2\Omega\eta
   + {\sin 2\Omega\eta\over 2\Omega}\right],\\
  \left<\right. Q_A,P_A\left.\right> =\left<\right. Q_B,P_B\left.\right>
    &\approx& {\hbar\gamma\over 2\Omega}\left[ \left(\coth
    {\pi\Omega\over a}-1\right) +  {2\over\pi}(\Lambda_0-\ln a)
    \sin 2\Omega\eta + \cos 2\Omega\eta\right],\\
  \left<\right. P_A^2\left.\right> =\left<\right. P_B^2\left.\right>
    &\approx& {\hbar\Omega\over 2} + \nonumber\\
    & & {\hbar\gamma\Omega}\left[
    \left(\coth {\pi\Omega\over a}-1\right)\eta+
    {2\over\pi\Omega}\left[ \Lambda_1-\ln a +
    (\Lambda_0-\ln a)\cos^2\Omega\eta\right]
    - {\sin 2\Omega\eta\over 2\Omega}\right],
\end{eqnarray}
one finds that, for the detectors initially in their ground states,
\begin{eqnarray}
a_{A+}=a_{B+}&\approx&{\Omega\over 2\hbar}+{\gamma\over\hbar}\left[{1\over\pi}
  \left[\Lambda_1-\ln a +(\Lambda_0-\ln a)\cos 2\Omega\eta \right] -
  {1\over 2}\sin 2\Omega\eta \right] ,\\
a_{A-}=a_{B-} &\approx& {\gamma\over\hbar}\left[ {2\Omega\eta\over
  e^{2\pi\Omega/a}-1} + {1\over\pi}(\Lambda_0+\Lambda_1-2\ln a)\right],\\
a_{X+} &\approx& {\gamma\over\hbar\sinh{\pi\Omega\over a}}\left[ 2\Omega\tau
  \cos 2\Omega\tau + \left( 1-{\pi\Omega\over a}\coth{\pi\Omega\over a}\right)
  \sin 2\Omega \tau\right],\\
b_{X+} &\approx& {\gamma\over\hbar\sinh{\pi\Omega\over a}}\left[ -2\Omega\tau
  \sin 2\Omega\tau + \left( 1-{\pi\Omega\over a}\coth{\pi\Omega\over a}\right)
  \cos 2\Omega \tau\right],
\end{eqnarray}
up to $O(\gamma)$, so that
\begin{eqnarray}
  \rho^R_{11,00} &\approx& -gK^2 {a_{X+}-i b_{X+} \over \left(
    {K^2\over 2}+ a_{A+}\right)^2}, \label{r1100WCL}\\
  \rho^R_{01,01} &\approx& -gK^2 {a_{B-} \over \left(
    {K^2\over 2}+ a_{A+}\right)^2}. \label{r0101WCL}
\end{eqnarray}
Now the growing rates of $|\rho^R_{11,00}|$ and $|\rho^R_{01,01}|$ at
large $\tau$ agree well with those estimated by TDPT$_\infty$, where
both $\eta$ and $\tau$ are replaced by infinities.

Nevertheless, one has to be careful in calculating the ratio of different
matrix elements in TDPT$_{\infty}$. Indeed, at $\Omega\tau\gg \Lambda_0$,
$\Lambda_1$, $|\rho^R_{01,01}|$ is approximately proportional to $\eta=
\tau-\tau_0$ while $|\rho^R_{11,00}|$ is approximately proportional to
$\tau$, so that $|\rho^R_{11,00}|/|\rho^R_{10,10}| \approx e^{\pi\Omega/a}
\times 2\tau/\eta$, where $\tau$ and $\eta$ do not simply cancel each other.
$|\rho^R_{11,00}|/|\rho^R_{10,10}|$ is actually time-varying, and Eq.(18)
in \cite{Reznik03} is correct only at the moment $\eta = 2\tau$, or 
$\tau=-\tau_0$, which is almost at the end of the TDPT$_\infty$ regime.

Note that, in weak coupling limit,
\begin{eqnarray}
  a_{X+} &\approx& -{\left<\right.Q_A,Q_B\left.\right>\over 8
    \left<\right.Q_A^2\left.\right>\left<\right.Q_B^2\left.\right>}
    + {1\over 2\hbar^2}\left<\right.P_A,P_B\left.\right>,\\
  b_{X+} &\approx& -{1\over 4\hbar}\left[
    {\left<\right.P_A,Q_B\left.\right>\over\left<\right.Q_B^2\left.\right>}+
    {\left<\right.Q_A,P_B\left.\right>\over\left<\right.Q_A^2\left.\right>}
    \right],\\
  a_{B-} &\approx& {1\over 8\left<\right.Q_B^2\left.\right>} -
    {1\over 2\hbar^2}\left<\right.P_B^2\left.\right>, \label{aBmWCL}
\end{eqnarray}
so the behavior of $\rho^R_{11,00}$ in $(\ref{r1100WCL})$ is highly
dependent on the behavior of the cross correlators, while $\rho^R_{01,01}$
in $(\ref{r0101WCL})$ is not.

Note also that $(\ref{truncRho})$ in ultraweak coupling limit is
numerically consistent with the TDPT results in Appendix \ref{TDPTapx}
in the time interval $\Omega^{-1}\ll \tau-\tau_0 \ll 1/\gamma$ for $0
< -\tau_0 \ll 1/2\gamma$. Here $\tau$ can be negative or greater than
$|\tau_0|$. From the three-stage behavior of the $|\rho^R_{11,00}|$
in Appendix \ref{TDPTapx}, which depends on the fiducial time
$\tau_0$, one can see that TDPT still keeps some of the non-Markovian
features, though the damping behavior ($\sim e^{-2\gamma\tau}$) has
been lost in the coupling constant expansion (cf. the integrands of
$(\ref{R1010TDPT})$-$(\ref{R1100TDPT})$ with $(\ref{QXdef})$). On the
other hand, the validity range of TDPT$_\infty$ is more restricted.
TDPT$_\infty$ is valid only when both $|\rho^R_{11,00}|$ and
$|\rho^R_{11,00}|$ are linearly growing, namely, only in the section
$\Omega^{-1}\ll \tau < |\tau_0| \ll 1/2\gamma$ of the time interval
$\Omega^{-1}\ll \tau-\tau_0 \ll 1/\gamma$ for TDPT. TDPT$_{\infty}$
results are Markovian because even the memory of the initial moment
is lost after the integration domain is extended.

\end{appendix}

\end{document}